\documentclass[acmlarge,anonymous=false,nonacm]{acmart}

\usepackage{amsmath}
\usepackage{algorithm}
\usepackage{algorithmic}
\usepackage{listings}
\usepackage{xcolor}
\usepackage{graphicx}
\usepackage{subcaption}
\usepackage{url}

\newcommand{\acronym}{\textit{X-BCD}}

\AtBeginDocument{%
  }

\begin{document}

\title{\acronym{}: Explainable Sensor-Based Behavioral Change Detection in Smart Home Environments}

\author{Gabriele Civitarese}

\email{gabriele.civitarese@unimi.it}
\author{Claudio Bettini}
\email{claudio.bettini@unimi.it}
\affiliation{%
  \institution{EveryWare Lab, Dept. of Computer Science, University of Milan}
  \city{Milan}
  \country{Italy}
}

\renewcommand{\shortauthors}{Civitarese and Bettini}

\begin{abstract}
  Behavioral changes in daily life activities at home can be digital markers of cognitive decline. However, such changes are difficult to assess through sporadic clinical visits and remain challenging to interpret from continuous in-home sensing data. Extensive work has been done in the ubiquitous computing area on recognizing activities in smart homes, but only limited efforts 
  have
  focused on analysing the evolution of patterns of activities, hence identifying behavior changes. 
  In particular, understanding how daily habits and routines evolve and reorganize (e.g., simplification, fragmentation) is still an open challenge for clinical monitoring and decision support. 
  
  In this paper, we present \acronym{}, an explainable, unsupervised framework for detecting and characterizing changes in activity routines from multimodal smart home sensor data, combining change point detection and cluster evolution tracking. 
  To support clinical interpretation, detected changes in routines are transformed into natural-language explanations grounded in interpretable features. Our preliminary evaluation on longitudinal data from real MCI patients shows that \acronym{} produces interpretable descriptions of behavioral change, as supported by cohort-level comparisons, expert assessment, and parameter sensitivity analysis. 
\end{abstract}



\received{20 February 2007}
\received[revised]{12 March 2009}
\received[accepted]{5 June 2009}

\maketitle

{\color{red}\bfseries This manuscript is currently under review and may be subject to revision.}

\section{Introduction}


As life expectancy continues to rise in modern societies, aging-related health challenges are becoming increasingly prominent. Digital technologies offer the opportunity to move beyond conventional healthcare routines, enabling scalable solutions for remote care and more effective management of health and independence among aging populations~\cite{chen2023digital}.
A common health issue that elderly subjects may face is Mild Cognitive Impairment (MCI): a pre-dementia stage where, while the subject keeps its core functional independence in daily activities, it is associated with a decline in cognitive abilities and/or executive functions, exceeding what would be expected for age and education~\cite{gauthier2006mild}. MCI can remain relatively stable or progress to dementia, depending on the underlying pathologies (e.g., neurodegenerative diseases). MCI affects between 12\% to 18\% of people aged 60 or older, and about one-third of people living with MCI due to a neurodegenerative disease may develop dementia within five years~\cite{alzheimer2022more}. 
Notably, an early indicator of transition from MCI to dementia is the functional decline in performing Activities of Daily Living (ADLs)~\cite{lussier2018early}. Indeed, subtle behavioral changes in the execution of ADLs (e.g., cooking, eating, personal care) can appear up to ten years before a diagnosis of dementia~\cite{peres2008natural}. However, MCI patients undergo relatively short and sporadic visits with medical experts who often can not accurately assess such decline.

In the last decades, many research groups have investigated how 
wearables and environmental sensors in smart homes can be used to continuously and unobtrusively recognize ADLs~\cite{babangida2022internet,liciotti2020sequential}.
The sensor-based recognition of ADLs performed by MCI subjects in their homes has already shown to provide potentially useful indicators of cognitive decline~\cite{riboni2016smartfaber}, but research on how to properly identify and quantify changes is still limited.  In the ADL recognition literature, activities are typically defined as short-term, observable, goal-oriented processes that can be formally described and directly inferred from sensor data (e.g., walking, cooking, eating). According to the survey in~\cite{arrotta2025multi}, activities are temporally bounded and semantically coherent, making them suitable as atomic units for recognition and evaluation. 

Unlike activities, \textit{behavior} refers to long-term patterns emerging from the aggregation and temporal evolution of recurring activities, capturing habitual routines, regularities, and gradual changes over extended periods. This distinction shifts the analytical focus from isolated activity recognition to the modeling of longitudinal trends and deviations that are more relevant for assessing functional change and well-being in smart home environments. Sensor-based Behavioral Change Detection (BCD) methods have been proposed to infer when behavioral shifts occur by computing a drift score that quantifies the magnitude of each detected change~\cite{sprint2020behavioral}. However, this aggregated measure offers little insight into how activity routines change. In MCI,  a behavioral change often reflects a reorganization of daily life, which may involve simplification, where task variability collapses to reduce cognitive load, fragmentation of previously coherent habits, loss of specific subroutines, or the development of compensatory behaviors to cope with emerging functional gaps~\cite{johansson2015cognitive}.  When a behavioral change is summarized solely through an overall drift score, these structurally and temporally meaningful reorganizations are averaged out. Understanding how behavioral patterns reorganize is therefore essential to improve the interpretability of monitoring systems and to provide clinically relevant information to support diagnosis.

In this paper, we propose \acronym{}, a novel approach to detect and characterize behavioral changes in smart home environments, generating natural language descriptions suitable for clinicians. \acronym{} aims at supporting clinicians in identifying interpretable digital markers that may indicate early symptoms of functional or cognitive decline. \acronym{} follows an ``\textit{interpretable-by-design}'' paradigm, proposing a novel integration of: (i) unsupervised change point detection to identify when behavioral changes occur; (ii) cluster evolution tracking to determine how routines reorganize for each detected change; and (iii) Large Language Models (LLMs) fine-tuned on medical data to summarize and explain observed behavioral reorganization in natural language. \acronym{} operates in a fully unsupervised setting, where labeled data are not available. In this context, interpretability is essential to enable clinicians to examine, contextualize, and validate detected behavioral changes.

While, in principle, it could be possible to obtain deep behavioral embeddings (e.g., using self-supervised approaches), they require large-scale datasets to learn stable and transferable representations and are difficult to interpret, thus posing additional challenges. For this reason, \acronym{} leverages handcrafted features grounded in domain knowledge and associated with clear semantic meaning. This results in a structured and interpretable feature space that supports clustering and change detection even under limited data availability, facilitating descriptive analysis of longitudinal behavioral change.

We conducted a preliminary evaluation of \acronym{} on a real longitudinal multi-modal dataset we collected in the homes of $17$ real MCI patients, covering multiple behavioral dimensions.
We show that \acronym{} produces interpretable descriptions of behavioral change, as supported by cohort-level comparisons, expert assessment, and parameter sensitivity analysis. 

The contributions of this paper are threefold:
\begin{itemize}
    \item We propose \acronym{}, a novel eXplainable Behavioral Change Detection approach for smart home environments, integrating change point detection and cluster evolution tracking to characterize behavioral change at a fine granularity, capturing how routines persist, drift, emerge, or disappear over time.
    \item \acronym{} generates natural-language descriptions of the detected behavioral changes understandable by clinicians. 
    \item We apply \acronym{} to real-world longitudinal data collected from individuals with MCI, illustrating how the proposed analysis reveals 
    different behavioral change patterns between the two cohorts of patients previously diagnosed by clinicians.
\end{itemize}

\section{Related work}

\subsection{Change Point Detection in Time Series}

Change Point Detection (CPD) aims to identify time points at which the statistical properties of a time series undergo significant changes, indicating transitions between different underlying regimes~\cite{liu2013change}. CPD has been widely adopted in ubiquitous computing and health-related applications, including sensor-based health monitoring and HAR~\cite{bosc2003automatic,staudacher2005new, gargoum2021limiting,aminikhanghahi2017using}. Unlike time-series anomaly detection, which focuses on isolated outliers or irregular events~\cite{zamanzadeh2024deep}, CPD targets sustained shifts in the data-generating process and is therefore more suitable for modeling long-term behavioral evolution. CPD methods can operate in supervised or unsupervised settings and in offline or online modes~\cite{aminikhanghahi2017survey}; however, in realistic health monitoring scenarios, annotated ground truth is rarely available, making unsupervised approaches particularly attractive. Offline methods analyze the full time series retrospectively to identify regime boundaries, whereas online methods aim to detect changes as early as possible, trading off detection delay for reliability.

Despite their effectiveness in segmenting time series, CPD methods typically operate at the level of statistical distributions or feature aggregates. As a result, detected change points are often summarized through scalar scores or abstract cost functions, offering limited insight into how complex human behaviors reorganize across segments. This limitation becomes particularly critical in clinical contexts, where interpretability and semantic grounding are essential.


\subsection{Sensor-Based Approaches for Cognitive Decline Assessment}

Smart home sensing technologies have been extensively investigated as a means to unobtrusively monitor cognitive health and functional abilities in older adults~\cite{dawadi2013automated, lussier2018early}.
Several studies frame cognitive decline detection as an anomaly detection problem, identifying deviations from a learned baseline of \textit{normal} behavior~\cite{hoque2015holmes, fahad2021activity, riboni2016smartfaber}. However, isolated anomalies may reflect noise, contextual factors, or short-term disruptions rather than persistent, clinically meaningful behavioral changes.


Other approaches leverage smart home sensor data to profile the inhabitants, to classify their cognitive status (e.g., MCI, Alzheimer's Disease, etc.)~\cite{khodabandehloo2021healthxai,tan2024predicting, teh2022predictive}. While valuable, these methods typically rely on relatively short observation windows and provide static snapshots of cognitive status, overlooking the progressive and individualized nature of cognitive decline.

Behavioral Change Detection (BCD) methods aim to identify persistent behavioral changes from sensor data~\cite{sprint2020behavioral}. Most existing approaches formulate BCD as a change point detection problem, where a quantitative score is computed to capture the magnitude of behavioral shifts over time. For example, Dynamo~\cite{prenkaj2023unsupervised} employs dynamic clustering to detect gradual behavioral changes and provides interpretability by highlighting the features that most contribute to each detected change. 
%
While effective in identifying when changes occur, such approaches tend to summarize behavioral evolution through aggregated scores. This abstraction often obscures the structure of behavioral reorganization, such as the emergence of new routines, the disappearance of established habits, or the fragmentation and merging of previously coherent patterns. In the context of cognitive impairment, these structural transformations are often more clinically informative than the mere magnitude of change~\cite{johansson2015cognitive}.

\subsection{Explainable AI in Sensor-Based Behavioral Monitoring}

Explainable AI (XAI) has highlighted the importance of making model outputs understandable and actionable for human stakeholders, particularly in 
safety- and health-critical domains~\cite{dwivedi2023explainable, miller2019explanation}. In sensor-based behavioral monitoring, explainability is essential not only to justify individual model decisions, but also to support longitudinal reasoning about how daily routines evolve over time. Existing approaches rely on latent representations or post-hoc explanations, providing explanations in terms of abstract features or importance scores~\cite{alharthi2025explainable, prenkaj2023unsupervised, sprint2020behavioral}.
These methods offer limited insight into the structural reorganization of everyday behavior. 
This motivates the need for explainable-by-design approaches that ground behavioral change detection in interpretable routine-level representations that can be directly mapped to meaningful activities.
Finally, while Large Language Models (LLMs) have been recently proposed to explain the output of ADLs recognition in smart homes~\cite{fiori2025leveraging}, their application to behavioral change detection is currently unexplored.

\subsection*{\textbf{Positioning of \acronym{}}}
In contrast to prior work, \acronym{} explicitly models behavioral change at the level of routines and their evolution over time. By integrating change point detection with cluster evolution tracking, the proposed approach captures how behavioral patterns persist, drift, emerge, or disappear across longitudinal segments. Furthermore, \acronym{} emphasizes explainability by design, transforming routine-level changes into natural-language descriptions grounded in interpretable features. This combination addresses a key gap in existing literature: the lack of methods that jointly support unsupervised behavioral change detection, structural interpretation of routine reorganization, and clinician-oriented explanations.




%


\section{\acronym{}: eXplainable Behavioral Change Detection}


\subsection{Problem Formulation}

\paragraph{Behavioral representation.}
\acronym{} is designed to operate under flexible behavioral representations and temporal granularities. 
Let $\mathcal{D} = \{D_1, D_2, \dots, D_m\}$
denote the set of behavioral dimensions of interest. To provide some examples, the coarsest representation would be $D = \{\textit{global behavior}\}$, an intermediate representation could be
$\mathcal{D} = \{\textit{sleep}, \textit{nutrition}, \textit{physical activity}, \textit{outdoor mobility}\}$,
and a finer-grained decomposition may look like
$\mathcal{D} = \{\textit{sleep duration and continuity}, \textit{sleep schedule}, \textit{sleep-stage composition}, \dots\}$.
Behavioral representations involve fundamental trade-offs. Coarse-grained behavioral representations are comprehensive but may obscure changes confined to specific behavioral components, which often reflect distinct physiological mechanisms and temporal dynamics. Fine-grained representations increase component-level sensitivity and interpretability, at the cost of higher computational complexity and a proliferation of detected change events, complicating downstream interpretation and prioritization. 

Let $g$ denote the temporal granularity at which behavioral data are modeled (e.g., hourly, daily, weekly). Very fine temporal resolutions (e.g., hourly) may emphasize short-term variability and sensor noise, biasing detection toward transient fluctuations and low-level actions while increasing computational burden. Conversely, overly coarse resolutions (e.g., monthly) average over time, potentially suppressing slow or subtle behavioral changes that remain clinically or scientifically relevant.

Hence, the choice of $\mathcal{D}$ and $g$ is determined by the available sensing infrastructure and by stakeholder objectives.

\paragraph{Sensing.}
From a sensing perspective, \acronym{} assumes a home environment equipped with environmental sensors (e.g., motion sensors, temperature and humidity sensors, plug sensors, sleep sensors) capturing the subject’s interaction with the environment, as well as one or more wearable devices (e.g., a smartwatch). Consistent with prior work, we assume either a single-occupant smart home or a sensing system capable of reliably attributing sensor activations to the monitored subject~\cite{arrotta2025multi}.

Let $\mathbf{S} = \{S_1, S_2, \dots, S_n\}$ denote the set of deployed sensors. For each behavioral dimension $D_i \in \mathcal{D}$, let $\mathbf{S}_{D_i} \subseteq \mathbf{S}$ be the subset of sensors whose signals 
can reveal behaviors in 
that dimension. For instance, the dimension \textit{indoor mobility} may be associated with a set of motion or presence sensors, enabling the estimation of time spent in different rooms and transitions between rooms. Sensor subsets associated with different dimensions are not required to be disjoint; in general, for $D_i \neq D_j$, it may hold that $\mathbf{S}_{D_i} \cap \mathbf{S}_{D_j} \neq \emptyset$.

\paragraph{Behavioral time series.}
\acronym{} analyzes each behavioral dimension of a subject separately. Given a dimension $D \in \mathcal{D}$, and a temporal granularity $g$, the signals generated by the sensors in $\mathbf{S}_{D}$ are aggregated according to $g$ to form a multivariate time series
\[
\mathbf{X}_{D} = \{\mathbf{x}_{D,t_1}, \mathbf{x}_{D,t_2}, \dots, \mathbf{x}_{D,t_k}\},
\]
where each $\mathbf{x}_{D,t} \in \mathbf{X}_{D}$ represents a feature vector compactly encoding behavioral dimension $D$ at the time granule with index $t$ based on raw data generated from $S_D$, and $k$ is the number of time steps induced by $g$ over the observation horizon. For instance, let $g=\textit{day}$, $D=\textit{nutrition}$ and $S_D=\{\textit{fridge door}, \textit{microwave}, \textit{stove}, \textit{kitchen PIR}\}$. In this scenario, each $\mathbf{x}_{D,t}$ is represented as a multidimensional vector of daily statistics inferred from $S_D$ at day $t$,
for example \textit{number of fridge openings}, \textit{fridge door-open duration},
\textit{microwave active duration}, \textit{electric stove active duration},
and \textit{kitchen presence duration}, among others.

\paragraph{Problem definition.}
The objective of \acronym{} is threefold. For each behavioral dimension \(D \in \mathcal{D}\):

\begin{enumerate}
\item \textbf{Behavioral change detection:} the goal is to identify a sequence of time indices
\[
\mathcal{T}_{D} = \{\tau_1,\tau_2,\dots,\tau_m\} \subseteq \{t_1,\dots,t_k\}, 
\qquad  \tau_1 < \tau_2 < \dots < \tau_m,
\]
at which the underlying behavioral process generating \(\mathbf{X}_{D}\) undergoes a statistically or behaviorally meaningful change. We call each of these indices a \textit{change point}.
Each change point may correspond to shifts in the distribution, dynamics, or structure of the multivariate time series, and is assumed to be sparse relative to the observation horizon.
Hence, the time series  is partitioned into the following time segments:  $$[t_1,\tau_1), [\tau_1, \tau_2), \dots, [\tau_{m-1},\tau_m), [\tau_m,t_{k}]$$

Intuitively, each segment represents a time interval in which the behavior in that dimension is internally consistent, while a behavioral change occurs 
in the consecutive segment.
With $\mathbf{X}^{D}_{[\tau_{j-1},\tau_j)}$ we denote the subsequence of the time series $X_D$ in the segment $[\tau_{j-1},\tau_j)$.

\item \textbf{Routine reorganization characterization:}
Given a change point $\tau_j \in \mathcal{T}_D$, the goal is to characterize whether and how behavioral routines reorganize between
$\mathbf{X}^{D}_{[\tau_{j-1},\tau_j)}$ and $\mathbf{X}^{D}_{[\tau_j,\tau_{j+1})}$.
By routine, we mean a way of behaving that recurs over time, exhibiting a recognizable pattern in what is done and how it is done.
For example, in the sleep domain, a subject may exhibit two distinct routines: during weekdays, an early bedtime followed by an early and regular wake-up time and, during weekends, a systematically later bedtime accompanied by a correspondingly later wake-up time.
If, after a change point, the weekend sleep routine aligns with the weekday one, resulting in a single, more uniform sleep routine across the week, the system should classify this reorganization as a \emph{merge} of the two previously distinct routines.

\item \textbf{Behavioral change explanation:} for each pair of consecutive segments
\(\mathbf{X}^{D}_{[\tau_{j-1},\tau_j)}\) and
\(\mathbf{X}^{D}_{[\tau_j,\tau_{j+1})}\),
the goal is to generate a human-interpretable, natural-language explanation describing the nature of the behavioral change in terms of whether and how behavioral routines reorganize. 
\end{enumerate}


We assume that \acronym{} operates in a \emph{batch} setting. That is, the full multivariate time series $\mathbf{X}_{D}$ over the observation horizon is available for analysis, and behavioral change points are detected retrospectively after data collection is complete. Extending the framework to an online or streaming setting is left for future work, as we discuss in Section~\ref{subsec:online}.


\subsection{Overall pipeline}

\begin{figure}[h!]
    \centering
    \includegraphics[width=1\linewidth]{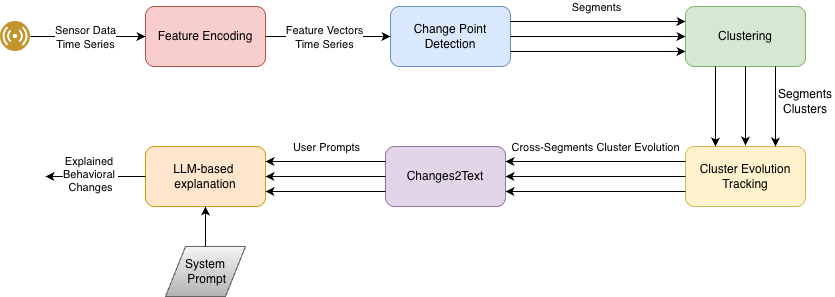}
    \caption{The high-level pipeline of \acronym{}. This pipeline is applied independently to each monitored behavioral dimension.}
    \label{fig:architecture}
\end{figure}

The high-level pipeline of \acronym{} is depicted in Figure~\ref{fig:architecture}. 
\acronym{} follows a modular, sequential pipeline designed to detect and interpret behavioral changes from longitudinal, multimodal sensor data in a transparent and clinically interpretable manner. The pipeline is applied independently to each behavioral dimension of interest.

The process begins with the \textsc{Feature Encoding} module, where raw sensor signals associated with a given behavioral dimension are aggregated at a fixed temporal granularity and transformed into interpretable, domain-specific behavioral features. These features capture meaningful behavioral aspects (e.g., sleep continuity, physical activity levels, or nutrition-related patterns) and include explicit temporal context variables to account for circadian and weekly regularities. The resulting multivariate time series provides a structured representation of behavior over time.

Next, the \textsc{Change Point Detection} module is in charge of running data-driven temporal segmentation. This step identifies time points at which the statistical properties of the behavioral feature distribution change, partitioning the time series into consecutive segments characterized by internally stable behavior. 

Within each detected segment, the \textsc{Clustering} module applies density-based clustering to group feature vectors that represent similar patterns. In this way, each cluster represents a \textbf{routine}: a behavior that occurs repeatedly within the segment.
Clusters are summarized through interpretable statistics, including their centroid, variability, and prevalence, which together characterize both the typical behavior and its consistency over time.

To analyze how behavior evolves, the \textsc{Cluster Evolution Tracking} module compares clusters across adjacent segments. This framework identifies different types of routine-level changes, including stable routines, gradual drifts, the emergence of novel routines, the disappearance of previously observed routines, and structural reorganizations such as routine splits or merges. This step provides a structured, routine-centric view of behavioral evolution, going beyond simple feature-level changes.

Then, the \textsc{Changes2Text} module transforms the output of the previous step into a natural language representation by using a fixed template. This textual representation encodes grounded, intermediate representations of behavioral change, including routine summaries and their evolution across segments.

This output is finally processed by the \textsc{LLM-based explanation} module, generating natural-language explanations that describe how behavioral patterns have changed, supporting clinicians in contextualizing detected changes, and assessing which changes may warrant further attention.


%
\subsection{Feature Encoding}
\label{subsec:features}
\label{subsubsec:interpfeat}

Given a behavioral dimension $D \in \mathcal{D}$, we assume the existence of a dimension-specific feature extraction function $f_g^D$ based on a fixed time granularity $g$ (e.g., hourly, daily, or weekly). Specifically, the multimodal time series $\mathbf{X}_D$ is first windowed into consecutive non-overlapping segments of duration $g$. For each time window indexed by $t$, the corresponding sensor observations $\mathbf{o}_{D,t}$ are summarized and mapped by $f_g^D$ to a feature vector:
\begin{equation}
    \mathbf{x}_{D,t} = f_g^D(\mathbf{o}_{D,t}) = [f_{D,1}, f_{D,2}, \ldots, f_{D,n}]^\top \in \mathbb{R}^n,
    \label{eq:feature_vector}
\end{equation}
where $\mathbf{o}_{D,t}$ denotes the observations collected from sensors $S_D$ during time window $t$, and $f_{D,j}$ corresponds to the $j$-th domain-specific behavioral feature.

For interpretability, $f_g^D$ is defined in a knowledge-driven manner by consulting domain experts (e.g., clinicians). This ensures that the most important behavioral aspects for stakeholders are captured. For example, when considering a daily time granularity ($g=\text{day}$), features in the sleep domain may include total sleep time, time spent in different sleep stages (REM, deep, and light), and the number of nocturnal awakenings. In the physical activity domain, representative features include daily step counts, physical activity intensity, and average heart rate.

Time-of-day features (e.g., sleep start/end times) are inherently periodic, defined on a circular domain rather than a linear one. For example, 11:00 pm and 01:00 am are only two hours apart, yet their raw numerical representations (11 and 1) would appear maximally distant in Euclidean space.  
To preserve this circular topology, each temporal variable \(h\) (measured in hours from midnight) is encoded using sine and cosine transformations:
\begin{equation}
    h_{\sin} = \sin\!\left(2\pi \frac{h}{24}\right), \qquad
    h_{\cos} = \cos\!\left(2\pi \frac{h}{24}\right).
    \label{eq:cyclic_encoding}
\end{equation}
This mapping embeds the one-dimensional cyclic variable onto the unit circle in \(\mathbb{R}^2\), ensuring that
\begin{equation}
    \|[h_{\sin}, h_{\cos}]^\top - [h'_{\sin}, h'_{\cos}]^\top\|_2
\end{equation}
reflects true temporal proximity modulo \(24\,\mathrm{h}\).  
Consequently, Euclidean distances between encoded times are consistent with their angular distance on the clock, and transitions across midnight (e.g., 23:00 → 01:00) are represented as smooth and continuous rather than as abrupt discontinuities.  
This encoding is therefore essential when clustering temporal behaviors that exhibit circadian periodicity. Importantly, this encoding preserves interpretability: each point on the unit circle corresponds uniquely to a clock time, so learned representations can be directly mapped back to human-readable times without loss of information.

Moreover, human behavior exhibits strong regularities governed by social and circadian structures (e.g., work schedules, leisure routines, recovery patterns).  
To capture these structured periodicities, we also include \emph{temporal context} features.
For instance, when $g=$\textit{day}, it is possible to introduce one-hot encoding for each weekday:
\[
\texttt{Monday}, \texttt{Tuesday}, \ldots, \texttt{Sunday} \in \{0,1\},
\]
and/or an aggregate weekend indicator
\[
\texttt{is\_weekend} = 
\begin{cases}
1, & \text{if day} \in \{\text{Saturday, Sunday}\},\\
0, & \text{otherwise.}
\end{cases}
\]
This representation allows the model to distinguish, for example, weekday and weekend differences in the activity patterns which are typical of humans.  Including explicit temporal context is essential for clustering and drift analysis because it often affects behaviors (e.g., later sleep onset and lower step count on weekends).

Each feature \(f_{D,j}\) is also associated with a semantic descriptor \(d_{D,j}\), which represents its natural language explanation. For example, with 
$g=$\textit{day}:
\begin{quote}
\texttt{light\_sleep\_duration}: ``The total time (in seconds) spent in light sleep stage during the night.''  
\end{quote}
\begin{quote}
\texttt{steps}: ``The total count of steps taken during the day.'' 
\end{quote}
As we will explain in Section~\ref{subsec: centroid_llm}, this mapping allows the model to produce interpretable textual summaries of routines when processed by the LLM in charge of generating natural language explanations.


Finally, features are standardized independently via z-score normalization:
\begin{equation}
    f'_{D,j} = \frac{f_{D,j} - \mu_{D,j}}{\sigma_{D,j}},
    \label{eq:standardization}
\end{equation}
where $\mu_{D,j}$ and $\sigma_{D,j}$ are the mean and standard deviation computed over all values of the features in the whole time-series $\mathbf{X}_{D}$. Without standardization, high-variance features would dominate both change point detection and clustering, potentially masking changes in relevant but lower-magnitude signals.
Global normalization is applied to preserve long-term comparability between segments.

\subsection{Data-Driven Segmentation via Change Point Detection}

Given a time series $\mathbf{X}_{D}$ of interpretable feature vectors, \acronym{} aims at finding \textit{behavioral change points} partitioning the time series into consecutive non-overlapping segments of at least $min\_size$ elements. Intuitively, we aim at detecting regime boundaries, that is, time intervals over which behavior can be considered statistically stable.

This is achieved with a data-driven Change Point Detection (CPD) approach, segmenting $\mathbf{X}_{D}$ based on intrinsic structural changes. The objective is to identify a set of change points $ \mathcal{T}_{D}^\star = \{\tau_1, \tau_2, \dots, \tau_m\}$ such that each segment  
\[
\mathbf{X}^D_{[\tau_{j-1},\tau_j)}
= [\mathbf{x}_{\tau_{j-1}}, \dots, \mathbf{x}_{\tau_j-1}]
\]  
exhibits homogeneous statistical properties, and its length is at least $min\_size$\footnote{Note that $\tau_j-1$ denotes the last time granule immediately preceding the change point $\tau_j$ (not to be confused with $\tau_{j-1}$, which is the previous change point). For example, if $g=\textit{day}$ and $\tau_j=\textit{April 20}$, then $\tau_j-1=\textit{April 19}$.}

. Segments are separated by points where significant structural shifts occur. Hence, this approach identifies sustained changes in the distribution of interpretable behavioral features, producing segments that are internally coherent and long enough to characterize routines. 
Technically, \acronym{} employs the Pruned Exact Linear Time (PELT) algorithm to detect change points~\cite{killick2012optimal}. The optimal segmentation is obtained by solving
\[
\mathcal{T}_{D}^{\star}
=
\arg\min_{\mathcal{T}_{D}}
\left\{
\sum_{j=1}^{m+1}
\mathrm{Cost}\!\left(\boldsymbol{X}^D_{[\tau_{j-1},\tau_j)}\right)
+
\beta\,m
\right\},
\qquad m = \lvert \mathcal{T}_{D} \rvert
\]
where $\mathcal{T}_{D}^{\star}$ is the optimal set of change points, $\mathrm{Cost}(\cdot)$ is a segment-wise cost function, $\beta$ is a penalty parameter controlling the number of detected change points, and $m$ is the number of change points of a solution $\mathcal{T}_{D}$ . For the sake of notation, assume that $\tau_0 = t_1$ and $\tau_{m+1}=t_k+1$, where $t_1$ and $t_k$ are the first and last time granules of $X_D$, respectively.
In our implementation, the cost function corresponds to the sum of squared $l_2$ residuals within each segment:  

\[
\mathrm{Cost}\!\left(\boldsymbol{X}_{[\tau_{j-1},\tau_j)}\right)
=
\begin{cases}
\displaystyle
\sum_{t=\tau_{j-1}}^{\tau_j-1}
\left\lVert
\boldsymbol{x}_t -
\overline{\boldsymbol{X}^D}_{[\tau_{j-1},\tau_j)}
\right\rVert_2^2,
& \text{if } \tau_j-\tau_{j-1} > \texttt{min\_size}, \\[10pt]
+\infty, & \text{otherwise}.
\end{cases}
\]
 
where $\overline{\boldsymbol{X}^D}_{[\tau_{j-1},\tau_j)}$ denotes the mean vector of the segment. The $l_2$ residual cost function inherently prioritizes the detection of acute feature changes, even if only a few features are involved. Because the cost function relies on the sum of \textit{squared} errors and all features are normalized before segmentation, outliers are penalized quadratically. Consequently, a singular acute change (e.g.,  a sharp increase in overnight bathroom visits) exerts a significantly stronger influence on the change point detection than minor noise distributed across all other features. 

The number of detected change points is influenced by the parameters $min\_size$ and $\beta$, which act on different aspects of the segmentation process.
The parameter $min\_size$ constrains the minimum allowable length of a segment: if chosen too small, the algorithm may over-segment the time series by producing short segments, possibly only separated because of noise, while excessively large values of $min\_size$ may prevent the detection of meaningful changes.
The penalty parameter $\beta$, instead, regulates the cost of introducing change points, balancing over-segmentation and under-segmentation so that detected change points correspond to sustained behavioral changes rather than isolated events.
In practice, $min\_size$ is typically set based on domain knowledge, whereas $\beta$ is tuned to control the overall complexity of the segmentation.


\subsection{Cluster Evolution Tracking}
\label{sec:clustering_drift}

In the following, we describe how
the segments identified by change point detection
are used in \acronym{}
to identify behavioral changes.

\subsubsection{Clustering} 
Starting from the first segment, \acronym{} sequentially analyses changes that occur in one segment (called \emph{Observation Period} or OP) with respect to the previous segment (called \emph{Reference Period} or RP).
The goal is to assess whether the behavioral dynamics observed during RP remain stable, drift, or reorganize in OP. Within each period, feature vectors are clustered to capture typical routines. By comparing these clusters across periods, we can quantify the degree of stability, transformation, or emergence of new routines.

\subsubsection{Cluster summary}

For each segment $\mathbf{X}_[{\tau_{j-1},\tau_j)}$, we apply a density-based clustering algorithm (e.g., DBSCAN or HDBSCAN) to obtain a set \(\mathcal{C}_{[\tau_{j-1},\tau_j)}\) of clusters. Intuitively, each cluster corresponds to a typical routine, since clustering identifies regions of the feature space that are densely populated by the data in the segment. Density-based clustering does not require predetermining the number of clusters (hence, the number of routines) in advance and, at the same time, allows noisy feature vectors to remain unassigned.

For each cluster $C \in \mathcal{C}_{[\tau_{j-1},\tau_j)}$, we compute:
\begin{itemize}
  \item \(|C|\) — cluster size (number of days in the cluster);
  \item \(\boldsymbol{\mu}_C \in\mathbb{R}^n\) — centroid (per-feature mean);
  \item \(\boldsymbol{\sigma}^2_C \in\mathbb{R}^n\) — per-feature variance;
  \item \(\mathrm{prop}_C = |C|/|\mathbf{X}_{[\tau_{j-1},\tau_j)}|\) — cluster proportion in the segment.
\end{itemize}

Intuitively, the centroid summarizes the routine captured by the cluster, the variance quantifies within-cluster heterogeneity, and the proportion indicates the prevalence of that pattern in the segment. 
%
In the following, we will refer to \(\mathcal{C}_{\mathrm{RP}}\) and \(\mathcal{C}_{\mathrm{OP}}\) for the sets of clusters in the reference period (RP) and observation period (OP), respectively.

\subsubsection{Similarity between clusters}

To compare clusters in $\mathcal{C}_{\mathrm{RP}}$ with the ones in $\mathcal{C}_{\mathrm{OP}}$, we need to define a similarity measure. Specifically, we combine two similarity metrics: one that considers the clusters' centroids and another that targets their variance.

Given two clusters \(C_i\) and \(C_j\), we compute their euclidean similarity-based distance:

\begin{equation} d_{ij} = |\boldsymbol{\mu}_{C_i} - \boldsymbol{\mu}_{C_j}|_2. \label{eq:euclidean_distance} \end{equation}

To normalize the centroid distance, we use their average standard deviation:
\begin{align}
\overline{\sigma^2}_{C} &= \frac{1}{n}\sum_{f=1}^{n} (\boldsymbol{\sigma}^2_C)_f,
\label{eq:meanvar}\\[4pt]
\bar{\sigma}_{ij} &= \sqrt{\frac{\overline{\sigma^2}_{C_i} + \overline{\sigma^2}_{C_j}}{2}}.
\label{eq:avgstd}
\end{align}

Finally, we map the normalized distance to a similarity score in $[0,1]$ using a Gaussian kernel:
\begin{equation}
\mathrm{Sim}_{\mathrm{centroid}}(C_i,C_j) =
\exp\left(-\frac{d{ij}^2}{2\bar{\sigma}_{ij}^2}\right)
\label{eq:centroid_sim}
\end{equation}
\textit{Intuition.} This formulation converts the Euclidean distance between cluster centroids into a dimensionless similarity score by normalizing it with a characteristic scale of the two clusters.
The quantity \(\bar{\sigma}_{ij}\) presents an aggregate measure of their internal spread, obtained by averaging feature-wise variances and pooling them across clusters. Dividing the centroid separation by this scale assesses whether the centroids are far apart relative to the typical size of the clusters themselves. As a result, the same absolute centroid distance is interpreted differently depending on cluster compactness: for tight clusters, even small separations are meaningful, whereas for diffuse clusters larger separations may be negligible. The Gaussian kernel then smoothly maps this normalized distance to $[0,1]$, yielding high similarity for centroids that are close compared to cluster spread and rapidly decreasing similarity otherwise.

Besides centroid similarity, we also compute per-feature min/max ratios and average them to obtain a variance-agreement score in $[0,1]$:
\begin{equation}
\mathrm{Sim}_{\mathrm{var}}(C_i,C_j)
=    \frac{1}{n}\sum_{f=1}^{n}
\frac{\min\!\big((\boldsymbol{\sigma}^2_{C_i})_f,(\boldsymbol{\sigma}^2_{C_j})_f\big)}
{\max\!\big((\boldsymbol{\sigma}^2_{C_i})_f,(\boldsymbol{\sigma}^2_{C_j})_f\big)}.
\label{eq:var_sim}
\end{equation}
\textit{Intuition.} This measure evaluates whether two clusters exhibit comparable internal dispersion patterns across features, independently of absolute scale. For each feature, the ratio between the smaller and larger variance provides a normalized agreement score equal to one when the variances coincide and decreasing smoothly toward zero as they diverge. Because the ratio is scale-invariant, it captures relative rather than absolute variability, preventing features with large numerical variance from dominating the comparison. Averaging these ratios across features yields a global variance-agreement score that is robust to isolated discrepancies and reflects whether the two clusters share similar heterogeneity profiles, under the assumption of feature-wise (diagonal) dispersion.

\paragraph{Combined cluster similarity.}
Weighted combination yields the final similarity score \(S_{ij}\in[0,1]\):
\begin{equation}
\mathcal{S}_{ij} = w_c\cdot\mathrm{Sim}_{\mathrm{centroid}}(C_i,C_j)
\;+\; w_v\cdot\mathrm{Sim}_{\mathrm{var}}(C_i,C_j),
\qquad w_c+w_v=1,
\label{eq:sim_score}
\end{equation}
with typical defaults \(w_c=0.7,\;w_v=0.3\).

The weights $w_c$ and $w_v$ control the relative importance of centroid proximity and variance agreement when comparing clusters across segments.
Increasing $w_c$ emphasizes changes of the average routine, while increasing $w_v$ emphasizes changes in behavioral variability and heterogeneity.
We prioritize centroid similarity ($w_c$ > $w_v$) because systematic shifts in routine-level behavior are more interpretable than changes in dispersion alone. Variance agreement moderates false matches caused by coincidental centroid proximity.

\subsubsection{Characterizing Behavioral Reorganization}

Our objective is to determine how clusters $\mathcal{C}_{\mathrm{RP}}$ from a reference period RP evolved into clusters $\mathcal{C}_{\mathrm{OP}}$ in the following observation period OP.

\begin{figure}[htbp]
    
  \centering
  \begin{subfigure}[t]{0.28\textwidth}
    \includegraphics[width=\linewidth]{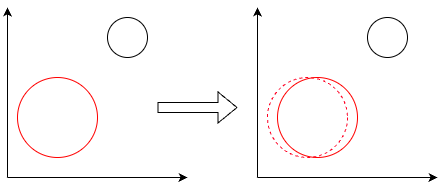}
    \caption{Stable}
  \end{subfigure}\hfill
  \begin{subfigure}[t]{0.28\textwidth}
    \includegraphics[width=\linewidth]{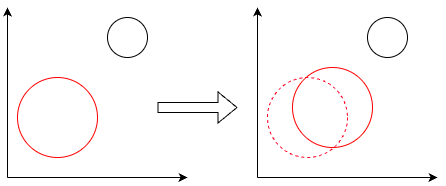}
    \caption{Drifting}
  \end{subfigure}\hfill
  \begin{subfigure}[t]{0.28\textwidth}
    \includegraphics[width=\linewidth]{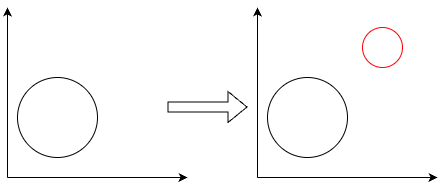}
    \caption{Novel}
  \end{subfigure}
  \vspace{6pt}

  \begin{subfigure}[t]{0.28\textwidth}
    \includegraphics[width=\linewidth]{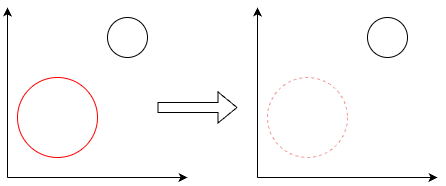}
    \caption{Disappeared}
  \end{subfigure}\hfill
  \begin{subfigure}[t]{0.28\textwidth}
    \includegraphics[width=\linewidth]{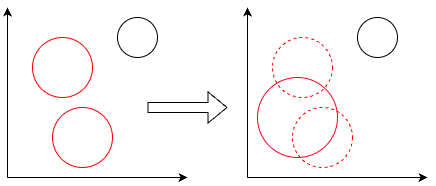}
    \caption{Merge}
  \end{subfigure}\hfill
  \begin{subfigure}[t]{0.28\textwidth}
    \includegraphics[width=\linewidth]{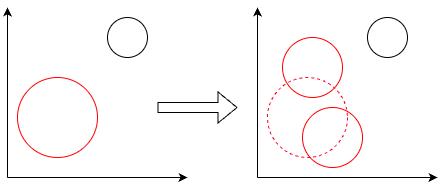}
    \caption{Split}
  \end{subfigure}

  \caption{Graphical intuition behind evolutionary continuity and topological changes.}
  \label{fig:clusterchanges}

\end{figure}

\acronym{} provides a granular classification of evolutionary changes inspired by cluster evolution literature~\cite{spiliopoulou2006monic}, distinguishing between two major classes of change (also depicted in Figure~\ref{fig:clusterchanges}):

\begin{enumerate}
    \item \textbf{Evolutionary Continuity:} This category tracks the direct descendant of a single cluster of RP in OP. We distinguish between \emph{stable} patterns, which maintain structural integrity with minimal property changes, and \emph{drifting} patterns, which persist but exhibit significant shifts in quantitative characteristics (e.g., centroid location, size, feature variance).
    \item \textbf{Topological Changes:} This category encompasses structural clusters reconfigurations that break one-to-one continuity:
    \begin{itemize}
        \item \emph{Novel Clusters:} Patterns that emerge in $\mathcal{C}_{\mathrm{OP}}$ with no significant antecedent in $\mathcal{C}_{\mathrm{RP}}$.
        \item \emph{Disappeared Clusters:} Patterns that were present in $\mathcal{C}_{\mathrm{RP}}$ but vanish without a significant successor in $\mathcal{C}_{\mathrm{OP}}$.
        \item \emph{Cluster Split:} A single cluster in $\mathcal{C}_{\mathrm{RP}}$ fragments into two or more distinct, significant clusters in $\mathcal{C}_{\mathrm{OP}}$.
        \item \emph{Merged Clusters:} Multiple clusters in $\mathcal{C}_{\mathrm{RP}}$ collapse into a single pattern in $\mathcal{C}_{\mathrm{OP}}$.
    \end{itemize}
\end{enumerate}

Following well-known approaches in the literature~\cite{spiliopoulou2006monic, oliveira2012framework}, we adopt a \textbf{thresholded bipartite adjacency check} to detect changes between clusters.
Each node in the bipartite graph represents a cluster, with clusters from RP on one side and clusters from OP on the other.
An edge is drawn between $C_i \in \text{RP}$ and $C_j \in \text{OP}$ if and only if $\mathcal{S}_{ij} \ge \tau_s$, where $\tau_s$ is a similarity threshold.
Then, the degree pattern of each node in this bipartite graph determines the type of change.
Specifically, the steps of our approach are the following:

\begin{enumerate}
  \item \textbf{Build Thresholded Directed Bipartite Graph:}
  Let $G=(V_{\mathrm{RP}},V_{\mathrm{OP}},E)$ be a bipartite \emph{directed} graph with
  \[
  V_{\mathrm{RP}}=\mathcal{C}_{\mathrm{RP}},\qquad V_{\mathrm{OP}}=\mathcal{C}_{\mathrm{OP}},\qquad
  E=\{(C_i,C_j)\mid S_{ij}\ge \tau_s\}\subseteq V_{\mathrm{RP}}\times V_{\mathrm{OP}}.
  \]
  Here $\tau_s$ is a similarity threshold.  Given $C_i\in V_{\mathrm{RP}}$ and $C_j\in V_{\mathrm{OP}}$, their neighborhoods are defined as:
  \[
  N^{+}(C_i)=\{\,C_j\in V_{\mathrm{OP}}\mid (C_i,C_j)\in E\,\},\qquad
  N^{-}(C_j)=\{\,C_i\in V_{\mathrm{RP}}\mid (C_i,C_j)\in E\,\}.
  \]

Here, $N^{+}(C_i)$ denotes the set of outgoing neighbors in the observation period (successors of $C_i$), while $N^{-}(C_j)$ denotes the set of incoming neighbors from the reference period (predecessors of $C_j$) in the directed bipartite graph.

 \item \textbf{Identify Evolutionary Continuity (One-to-One):}
  An edge $(C_i,C_j)\in E$ is a \textbf{continuity link} if $|N^{+}(C_i)|=|N^{-}(C_j)|=1$.
  Only for these one-to-one matches we compute a \textbf{drift score} $D_{ij}$ to label them as \emph{stable} or \emph{drifting} (see Section~\ref{subsec:driftscore}). 

\item \textbf{Identify Disappeared and Novel Clusters:}
  \begin{itemize}
    \item $C_i\in V_{\mathrm{RP}}$ is \textbf{disappeared} if $|N^{+}(C_i)|=0$ (no outgoing edge).
    \item $C_j\in V_{\mathrm{OP}}$ is \textbf{novel} if $|N^{-}(C_j)|=0$ (no incoming edge).
  \end{itemize}

  \item \textbf{Identify Cluster Splits (One-to-Many):}
  $C_i\in V_{\mathrm{RP}}$ is a \textbf{split} if $|N^{+}(C_i)|>1$; then $C_i$ splits into the successors $N^{+}(C_i)$.

  \item \textbf{Identify Cluster Merges (Many-to-One):}
  $C_j\in V_{\mathrm{OP}}$ is a \textbf{merge} if $|N^{-}(C_j)|>1$; then the predecessors $N^{-}(C_j)$ merge into $C_j$.

\end{enumerate}

\subsubsection{Drift score.}
\label{subsec:driftscore}

Let $(C_i \in \mathcal{C}_{\mathrm{RP}}, C_j \in \mathcal{C}_{\mathrm{OP}})$ be a one-to-one link identified by the algorithm above. \acronym{} computes a \textit{drift score} $D_{ij}$ to quantify how much the cluster $C_j$ differs from $C_i$. 
When $D_{ij}$ is lower than a drift threshold $\tau_d$, \acronym{} classifies the one-to-one link as \textit{stable} (i.e., the routine remained stable from RP to OP), otherwise it is classified as \textit{drift} (i.e., the routine in RP exhibits a significant change in OP).
If $\tau_d$ is too low, minor fluctuations may be classified as drift, reducing specificity. On the other hand,  if $\tau_d$ is too high, meaningful but moderate routine changes may be labeled as stable.

Specifically, the score $D_{ij}$ is computed via three interpretable components:

\begin{itemize}

\item \textbf{Centroid drift:}

\begin{equation}C_\Delta =  1 - \mathrm{Sim}_{\mathrm{centroid}}(C_i,C_j) \end{equation}

\item \textbf{Proportion change:}\quad

\begin{equation}P_\Delta =|\mathrm{prop}_{C_j}-\mathrm{prop}_{C_i}|\label{eq:prop_change}\end{equation}


\item \textbf{Variance change:}\quad

\begin{equation}V_\Delta =\frac{|\overline{\sigma^2}_{C_j}-\overline{\sigma^2}_{C_i}|}{\max(\overline{\sigma^2}_{C_i},\overline{\sigma^2}_{C_j})}\label{eq:var_change}\end{equation}

\end{itemize}

\paragraph{Composite drift score.}Combine the three components into a drift score $D_{ij}\in[0,1]$:\begin{equation}D_{ij} = \alpha_p P_\Delta +; \alpha_c C_\Delta + \alpha_v V_\Delta,\qquad \alpha_p+\alpha_c+\alpha_v = 1,\label{eq:drift_score}\end{equation}

Intuitively, proportion captures prevalence shifts (often clinically important), centroid drift captures systematic change in the pattern (spatial shift), and variance captures heterogeneity changes.
The weights control the contribution of each metric to the overall drift score. Emphasizing proportion change highlights shifts in routine prevalence, while emphasizing centroid drift captures systematic changes in behavioral characteristics. Variance change captures alterations in routine consistency.

In practice, these weights should be chosen to reflect domain priorities (e.g., defined by clinicians). A possibility is to assign greater importance to proportion and centroid shifts, which are easier to interpret and more likely to reflect meaningful behavioral reorganization. 
By default, \acronym{} adopts $\alpha_p = \alpha_c = \alpha_v = 0.33$, thus giving equal importance to each component.

\subsection{LLM-based behavioral changes interpretation}

For each change point, \acronym{} transforms the output of Clusters Evolution Tracking in natural language, employing a large language model (LLM) to explain behavioral reorganization. 


Explanations are intended for clinicians. Therefore, \acronym{} leverages an LLM fine-tuned on medical data to ensure the use of appropriate clinical terminology. Moreover, explanations should not include terms specific to the technique used to identify the changes (e.g., clusters, patterns, drift).
Unfortunately, this is what happens if we directly provide as input to the LLM the centroids' raw data and the output of cluster evolution tracking, no matter what prompt instructions we provide. 

For this reason, \acronym{} first builds a high-level representation of routines using the information on cluster centroids. The final explanation provided to clinicians is built by an LLM using the high-level representation.

\subsubsection{Transforming cluster centroids in textual descriptions}
\label{subsec: centroid_llm}

Given a change point, \acronym{} uses an LLM to convert each cluster centroid from RP and OP into a high-level textual representation of the routine it represents. The LLM is prompted with the following input:

\begin{itemize}
    \item The feature values of the centroid in their original space;
    \item Feature descriptors mapping sensor-derived variables (e.g., steps, sleep stages, activity intensity) to natural language explanations as descrived in Section~\ref{subsubsec:interpfeat};
    \item The number of elements within the period of occurrence;
    \item The percentage of the total data that it covers within the period it belongs (RP/OP).
    
\end{itemize}

The instructions require the use of plain natural language to describe the routine, avoiding inferring information not explicitly present in numerical data, and keeping a short description, avoiding redundancy.

Output example: \textit{``This patient's typical sleep routine involves spending approximately 8 hours (28807.5 seconds) asleep, distributed across light sleep (3.66 hours), REM sleep (1.77 hours), and deep sleep (4.34 hours). They typically wake up around 3 times during the night, with a total awake time of about 44.6 minutes (2677.5 seconds), and get out of bed roughly once. The total time spent in bed is about 8.75 hours (31485 seconds). The sleep period usually starts around 11:03 PM and ends around 7:48 AM. This routine does not occur on weekends.''}

\subsubsection{LLM interpretation of behavioral changes}

Once we obtain the description of the routine that each centroid represents, by using the information from the bipartite graph mapping clusters in RP to clusters in OP (including stable/drift classification for one-to-one links), we can prompt the LLM to explain and interpret the output of behavioral change detection.
%

For instance, considering drifted routines, the template is the following:

\begin{lstlisting}
In the reference period, the routine (proportion X%, number of days Y) is originally described as:

[NATURAL LANGUAGE DESCRIPTION OF ROUTINE IN RP]

In the observation period, this routine changed (now proportion W%, number of days Z), and it is  described as:

[NATURAL LANGUAGE DESCRIPTION OF CORRESPONDING DRIFTED ROUTINE IN OP]
\end{lstlisting}

As another example, the template for disappeared routines is the following:

\begin{lstlisting}
In the reference period (proportion X%, number of days Y), the routine was as:

[NATURAL LANGUAGE DESCRIPTION OF ROUTINE IN RP]

This routine is no longer observable in the observation period.
\end{lstlisting}

The LLM is hence prompted to produce the following assessment:
\begin{enumerate}
    \item \textbf{\textit{Global Behavioral Trend}}: a short summary of how routines changed overall.
    \item \textbf{\textit{Habit Dynamics}}: an objective and more detailed description of how each routine evolved across periods, focusing on net functional changes (e.g., gain, loss, compensation).
    \item \textbf{\textit{Potential Implications}}: hypothesizing whether the observed changes are consistent with resilience/compensation, simplification, or dysregulation of routines. 
\end{enumerate}




Note that the assessment of potential implications 
is not meant as a diagnostic tool, and the LLM is instructed to use cautious language, avoiding drawing definitive conclusions about the patient's health status.


In the following, we show an example of output ($D=$ \textit{physical activity}, $g=$day):

\begin{lstlisting}
A) Global Behavioral Trend:

The overall daily activity pattern shows a reduction in step count and elevation change. The duration of soft and moderate activity has decreased, while intense activity remains absent in both periods. The average heart rate has slightly increased.


B) Habit Dynamics:

The most frequent weekday routine observed in the reference period (around 69% of days) is no longer present. The Sunday routine has been reconfigured, showing a decrease in steps (from ~6700 to ~4100) and elevation change (from ~18 to ~13 steps), with a slight increase in average heart rate (from ~80 to ~82 bpm). Three new routines have emerged: a common weekday routine (~23% of days, ~3900 steps), a Saturday routine (~8% of days, ~2300 steps), and another weekday routine (~15% of days, ~2500 steps). These new routines generally involve fewer steps and less elevation change compared to the lost weekday routine, although the duration of soft and moderate activity varies.


C) Potential Implications:

The observed patterns, including the disappearance of a dominant weekday routine and the emergence of multiple new routines with lower functional output (steps, elevation), might reflect a simplification or fragmentation of the daily activity structure. The reconfiguration of the Sunday routine with reduced steps could be consistent with a shift in activity levels or duration on that specific day. These changes may suggest alterations in the patient's ability to maintain or execute established habits, or a change in the way they structure their daily activity.
\end{lstlisting}


\section{Experimental Evaluation}
\subsection{Dataset}

We evaluate \acronym{} on a dataset we collected within a longitudinal study conducted in collaboration with clinicians (i.e., neurologists and neuropsychologists) of a major hospital in Milan~\footnote{\url{https://ecare.unimi.it/it/pilots/serenade}}. To our knowledge, this is one of the largest datasets collected from the real homes of MCI patients, since it includes data from multiple modern sensing devices (smartwatch, sleep monitors, ...) and environmental sensors (smartplugs, temp\&humidity, open/close, presence, ...) continuously acquired for twelve to thirty months, depending on the patient. 
This data collection campaign has been approved by the hospital IRB, and all patients provided written informed consent. 
The dataset is pseudonymized (i.e., it has been curated at the acquisition time to mitigate the risk of revealing the patients' identities to data analysts).
The study involved 19 patients, but two have been excluded for various reasons. The remaining 17 are subdivided into two cohorts:

\begin{itemize}
    \item \textbf{D}: subjects diagnosed with mild cognitive impairment (MCI) attributable to an underlying neurodegenerative disease (8 subjects).
    \item \textbf{ND}: subjects diagnosed with mild cognitive impairment (MCI) not attributable to a neurodegenerative disease (9 subjects).
\end{itemize}

The assignment of each subject to the corresponding cohort was determined by clinicians before data collection, based on RMI imaging, neuropsychological tests, and other clinical evaluations.
Clinicians expect subjects in group \textbf{D} to progress relatively rapidly toward dementia, whereas subjects in group \textbf{ND} are expected to exhibit a slower rate of cognitive decline. Patients have been clinically re-evaluated every six-months.

In Section~\ref{subsec:considered_dimensions}, we describe in detail the set of sensing devices we considered in the experiments of this paper, while the detailed sensing infrastructure used for data collection can be found in~\cite{civitarese2025serenade}.
For each subject, the collected longitudinal data spanning multiple behavioral dimensions, including cooking habits, physical activity, sleep, indoor and outdoor mobility, and personal hygiene.
Since changes in MCI subjects are gradual and observable over long periods, the extended observation period of this dataset enables the study of meaningful behavioral changes.

We did not select alternative or additional datasets after exploring those commonly adopted in behavioral change detection, as they are predominantly synthetic or private; moreover, when publicly available, their data collection span is too short to support the identification of meaningful behavioral change (e.g., often limited to approximately one month)~\cite{prenkaj2023unsupervised,teh2022predictive,sprint2020behavioral}. 

\subsection{Evaluation Strategy}

Since the considered dataset does not provide ground truth annotations about the activities being performed or about when and how behavioral changes occurred, we can not evaluate \acronym{} in terms of change detection accuracy. Instead, we assess its ability to induce meaningful insights through cohort-level comparisons. Specifically, we analyze differences in change-point distributions and patterns of routine reorganization between neurodegenerative (\textbf{D}) and non-neurodegenerative (\textbf{ND}) cohorts. These observations are discussed, when appropriate, in relation to existing clinical literature on MCI, strictly as descriptive and hypothesis-generating evidence, without asserting diagnostic, prognostic, or individual-level validity. We do not compare \acronym{} against state-of-the-art methods, since (to the best of our knowledge) this work is the first to explicitly characterize routine evolution as its core contribution.

\subsection{Experimental Setup}

\subsubsection{Considered Dimensions}
\label{subsec:considered_dimensions}

For this study, we focus on a subset of the behavioral dimensions available from the dataset; they were selected in consultation with clinicians as the most promising for discriminating among the cohorts. 

\paragraph{Sleep}

Sleep disturbances are prevalent in MCI subjects, and changes are often linked to accelerated cognitive deterioration. Sleep was monitored using the \textit{Withings Sleep Analyzer}.\footnote{\url{https://www.withings.com/eu/en/sleep-analyzer}}. This device is an unobtrusive thin smart mattress collecting detailed information about sleep, including duration of sleep episodes, time spent in sleep stages, the number of times the subject wakes up and/or gets out of bed, heart rate variability while sleeping, snoring episodes, etcetera. We focused the evaluation of \acronym{} on the following sleep subdimensions, separately: 

\begin{itemize}
    \item \textbf{Sleep stages:} total time spent in REM, deep, light phases.
    \item \textbf{Sleep continuity and duration:} overall sleep time, number of wake-ups, duration of wakeup times, number of times the subject was out bed, total time in bed.
    \item \textbf{Sleep schedule:} times at which the person goes to bed and wakes up.
\end{itemize}

\paragraph{Physical Activity}

In MCI subjects, low activity levels are often associated with faster cognitive decline and a higher risk of progression to dementia. We monitor physical asking the subjects to wear a \textit{ScanWatch 2} smartwatch.\footnote{\url{https://www.withings.com/eu/en/scanwatch-2}}. The design of this smartwatch resembles the one of a normal watch, which helped us in making sure it was accepted by the participants. We focused the evaluation of \acronym{} on the following physical activity subdimensions, separately: 

\begin{itemize}
\item \textbf{Steps and elevation}:  number of steps, distance, elevation (i.e., stairs climbed, estimated by the device from altimeter-based altitude changes during physical activity).
\item \textbf{Physical Intensity}:  distribution of light, moderate, and intense activity (estimated by the device leveraging inertial and heart data).
\end{itemize}

\paragraph{Cooking}

Difficulties in carrying out Instrumental Activities of Daily Living (IADLs) are often considered an early indicator of functional decline and progression toward dementia. Cooking is a particularly representative IADL, as it integrates planning, memory, attention, and executive functions that are vulnerable in the earliest stages of cognitive impairment. While the homes in the dataset were equipped with multiple environmental sensors, for the sake of this work, we focused on \textit{cooktop usage} as a proxy for cooking.

In all participants' homes, kitchens were equipped exclusively with gas cooktops. Therefore, we used a temperature sensor placed above the cooktop to detect its usage. A preliminary analysis 
including diverse cooking tests showed that this sensing strategy produces characteristic temperature peaks corresponding to cooking events. We observed the same peak patterns in the dataset collected from participants' homes, occurring at times consistent with typical lunch preparation periods. Based on these observations, we applied peak-detection methods to isolate these cooking events and other methods to estimate their start and end times.

For the sake of behavioral analysis, we computed, for each time-of-day segment (morning, afternoon, evening, and night), the total cooktop usage duration, the number of cooking events, and the fraction of cooktop usage relative to the total daily usage.

\subsubsection{Adopted temporal granularity}

In our experiments, all behavioral dimensions are modeled at a daily granularity (i.e., $g=\textit{day}$). This choice represents a trade-off between finer-grained representations, more sensitive to short-term fluctuations and noise, and coarser temporal aggregations, which may obscure meaningful routine-level structure. Daily aggregation is also widely adopted in the literature for modeling longitudinal behavioral patterns~\cite{dawadi2016modeling}.
From a clinical perspective, daily representations preserve the natural variability of human routines while retaining sensitivity to changes in circadian patterns, which have been associated with cognitive decline~\cite{tranah2011circadian}.
Besides dimension-specific features, our daily representation also includes temporal features to distinguish weekdays from weekends, as described in Section~\ref{subsec:features}.

\subsubsection{Cluster Evolution Metrics}
\label{subsec:metrics}

To quantitatively assess how daily routines evolve across a change point, we analyze the output of cluster evolution tracking. For each type of change (e.g., drift, novelty), we compute the associated mass, defined as the amount of data involved in that change. This measure captures the relative impact of each evolution event. Intuitively, a drifting cluster spanning a small fraction of days contributes marginally to the drifting mass, whereas a cluster spanning a large fraction of days contributes proportionally more.
For \emph{stable}, \emph{drifted}, \emph{disappeared}, \emph{split}, and \emph{merged} change types, their mass is computed considering their clusters' proportion with respect to the reference period. On the other hand, novel clusters represent an asymmetric case, since they only appear in the observation period.

Formally, let $\mathcal{C}_{RP}$ and $\mathcal{C}_{OP}$ denote the sets of clusters in the reference period (RP) and observation period (OP), respectively. Let $|C|$ denote the number of data points assigned to cluster $C$, $|\mathcal{C}_{RP}|$ and $|\mathcal{C}_{OP}|$ be the total number of data points in the corresponding periods, and $|RP|$ and $|OP|$ the number of data points in the RP and OP.  Let $\mathcal{E} = \{\text{stable},\text{drifted},\text{novel},\text{disappeared},\text{split},\text{merged}\}$ 
denote the set of change types, and let $\mathcal{C}_e$ be the set of clusters classified as change type $e \in \mathcal{E}$. The \textit{overall mass} associated with change event type $e$ is defined as:
 
\[
M_e =
\sum_{C \in (\mathcal{C}_e \cap \mathcal{C}_P)}
\frac{|C|}{|P|},
\qquad
P =
\begin{cases}
RP, & \text{if } e \neq \text{novel},\\
OP, & \text{if } e = \text{novel}.
\end{cases}
\]

Hence, for each change point we compute the \emph{mass vector}:
\[
\mathbf{M} =
\bigl(
M_{\text{stable}},
M_{\text{drifted}},
M_{\text{novel}},
M_{\text{disappeared}},
M_{\text{split}},
M_{\text{merged}}
\bigr),
\]
which summarizes the proportion of data involved in each type of structural change.

As proposed in the literature, a unified change score may also be useful to summarize the overall magnitude of change. Following this idea, we also define the \emph{unstable routine mass} as the total mass associated with all non-stable routine changes. Let
$\mathcal{E}_{\text{unstable}} =
\mathcal{E} \setminus \{\textit{stable}\}$
The unstable routine mass is defined as
\[
M_{\text{unstable}} = \sum_{e \in \mathcal{E}_{\text{unstable}}} M_e
\]
This scalar quantity provides a compact summary of the overall magnitude of routine change across the detected change point.

\subsubsection{Implementation and parameter choice}

We implemented a prototype of \acronym{} in Python. 

For change point detection, we employed the well-established \textit{ruptures} library.~\footnote{\url{https://centre-borelli.github.io/ruptures-docs/}}
Based on clinicians suggestion, we opted for $min\_size=90$ to generate segments that last at least $3$ months, to capture changes that persist in time.
We tuned $\beta$ and the cluster evolution main parameters by running a sensitivity analysis, which we will describe in Section~\ref{sec:sensitivity}. Specifically, we opted for reasonable parameters located in a plateau region of the sensitivity analysis, where the output remains approximately constant under small perturbations for all the dimensions. Thus, we used $\tau_s = 0.5$ for the similarity threshold, $\tau_d = 0.25$ for the drift score threshold, and $\beta = 60$ for the penalty value of change point detection.

For clustering, we employed DBSCAN with cosine similarity. For each behavioral dimension, parameters were selected via a bounded grid search optimizing the silhouette score computed on clustered (non-noise) samples. We discarded configurations producing a single cluster or excessive noise. We fixed $min\_samples=5$ across all dimensions and observed modest variation in $eps$ (ranging from $0.23$ to $0.35$), reflecting differences in feature-space geometry across dimensions.

Considering LLM, due to privacy constraints, the use of third-party models was not permitted. Consequently, we adopted the MedGemma 27B model~\footnote{\url{https://research.google/blog/medgemma-our-most-capable-open-models-for-health-ai-development/}}
, deployed locally via Ollama on a high-performance, GPU-equipped server in our laboratory. This setup enabled LLM inference without external data transmission. Although we evaluated several alternative open-weight models, MedGemma consistently produced higher-quality explanations, likely because it is fine-tuned on medical-domain data.

\subsection{Results}

\subsubsection{Differences in detected change points}

\begin{figure}[h!]
    \centering
    \includegraphics[width=0.8\linewidth]{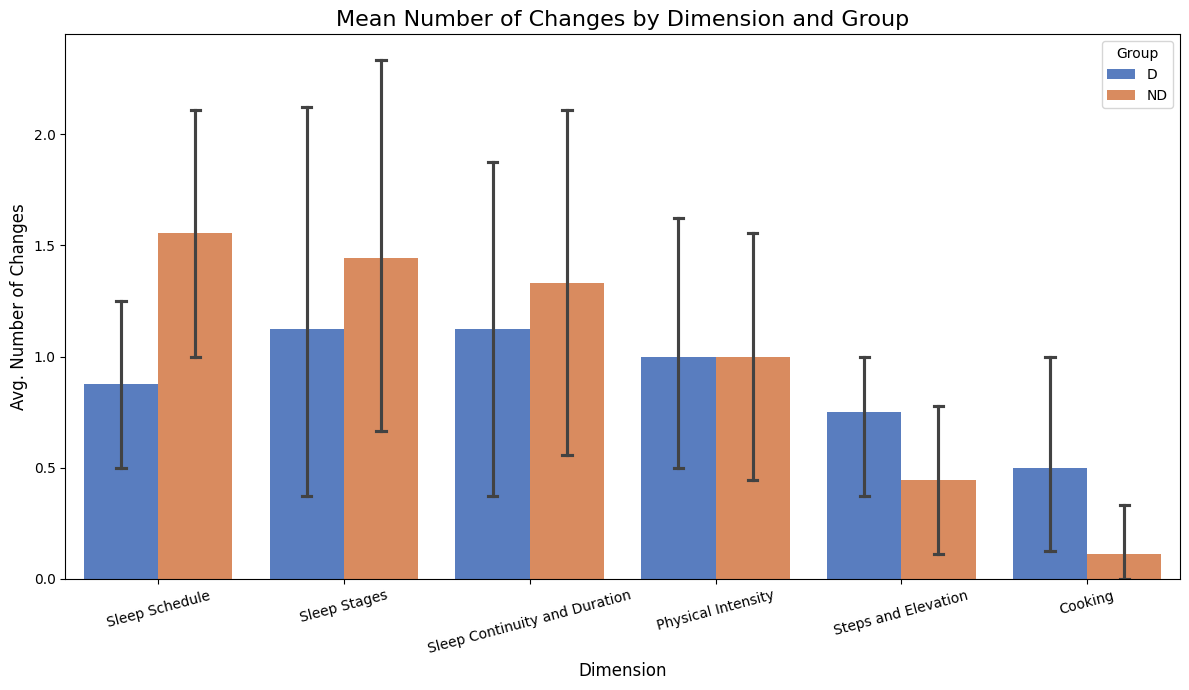}
    \caption{Difference in number of changes across cohorts for each dimension}
    \label{fig:num_change_diff}
\end{figure}

In Figure~\ref{fig:num_change_diff} we show how the two cohorts \textbf{D} and \textbf{ND} differ, on each dimension, in the number of detected change points.
Considering sleep dimensions, it is well-established that sleep disturbances are frequent in individuals with MCI and often co-occur with anxious and depressive symptoms, indicating a complex interplay between neuropsychiatric symptoms and sleep disruption in MCI populations~\cite{rozzini2009anxiety,song2021sleep}. We observed that group \textbf{ND} is associated with a higher number of detected changes compared to group \textbf{D}. Rather than indicating greater dysfunction, the higher number of detected changes in group \textbf{ND} likely reflects a better adaptive variability, whereas the relative stability observed in group \textbf{D} may represent a loss of flexibility imposed by neurodegeneration. This is consistent with the literature on MCI, since neurodegenerative processes are characterized by early synaptic dysfunction and loss, which limit flexibility and adaptive reorganization mechanisms~\cite{selkoe2002alzheimer}.

Considering steps and elevation, we observed a higher number of detected changes in group \textbf{D}, while few alterations were observed in group \textbf{ND}. Physical activity is strongly shaped by external constraints and daily routines, and in neurodegenerative MCI, progressive disruption of internal regulatory mechanisms may necessitate reorganization of activity patterns to maintain engagement. In contrast, in non-neurodegenerative MCI, preserved regulatory control may support stable activity routines, resulting in fewer detectable reorganizations despite cognitive impairment.

Notably, on the cooking dimension, we observed a single subject in group \textbf{ND} showing changes, compared to group \textbf{D}, where multiple subjects exhibited such changes. While these numbers are clearly small and not statistically relevant, they may be consistent with evidence in the existing literature. As mentioned before, cooking is a complex instrumental activity of daily living (IADL) well known to be particularly vulnerable in neurodegenerative forms of MCI. In group \textbf{D}, the gradual loss of coordination may lead individuals to reorganize when and how such activities are performed, whereas in group \textbf{ND}, where IADL-related cognitive systems are largely preserved, cooking behavior tends to remain more stable and requires less reorganization~\cite{aretouli2010everyday}.

Finally, changes in physical activity intensity did not show differences between the two groups, suggesting that this dimension is less sensitive to the distinction between neurodegenerative and non-neurodegenerative MCI, and may instead reflect general behavioral adjustments or externally driven factors that are common across MCI subjects, regardless of underlying etiology.

\subsubsection{Differences in change magnitude}

\begin{figure}[h!]
    \centering
    \includegraphics[width=1.0\linewidth]{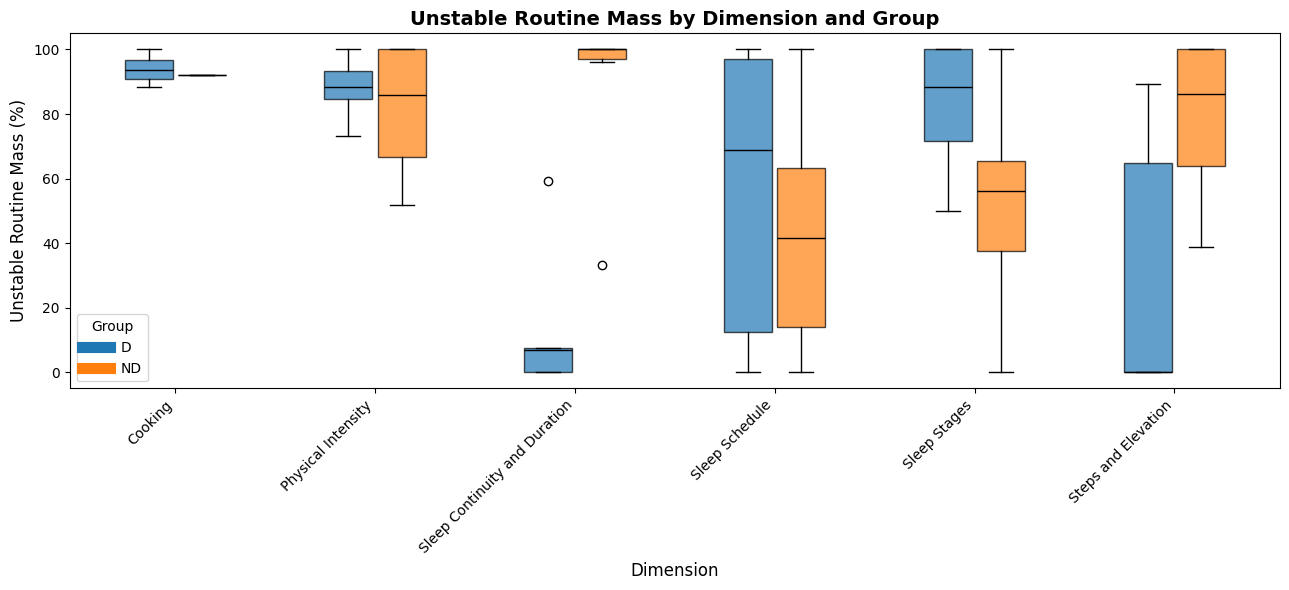}
    \caption{Distribution of unstable routine mass (i.e., drift, split, merge, novel, and disappeared) across cohorts for each dimension}
    \label{fig:mass_diff}
\end{figure}

To compare the two cohorts in terms of change magnitude, we first study the distribution of \textit{unstable routine mass} defined in Section~\ref{subsec:metrics}. This helps in understanding which dimensions deserve further analysis to characterize \textbf{D} and \textbf{ND} groups.
Figure~\ref{fig:mass_diff} shows how the unstable routine mass distributes for each dimension.

Among the sleep dimensions, it emerges that sleep stage reconfigurations are most predominant in the \textbf{D} group. This is consistent with existing literature, where sleep stages reconfiguration (e.g., reduced REM and hence less restorative sleep) is a known biomarker for the Alzheimer Disease~\cite{liguori2020sleep}. Surprisingly, the magnitude of change in sleep continuity and duration is significantly higher in the \textbf{ND} group and close to 0 in the \textbf{D} group. This may be because subjects in group \textbf{ND} may preserve more affective symptoms (e.g., anxiety and depression), which increase physiological and cognitive hyperarousal and makes sleep more reactive and easily interrupted. In contrast, neurodegenerative MCI is more often associated with apathy and reduced emotional reactivity~\cite{velayudhan2023apathy}, so sleep may appear less fragmented even though its microstructure and restorative quality (i.e., sleep stages) are impaired. 
Considering sleep schedule, the difference is less nuanced, even though it appears that group \textbf{D} is associated with more significant routine changes. In the literature, a low day-to-day stability in the circadian rhythm often anticipates conversion to dementia~\cite{li2020circadian}. Considering steps and elevation, group \textbf{ND} is associated with a higher unstable routine mass. This is in contrast with the average number of change points in this dimension, which was higher in group \textbf{D}. Clinically, this pattern is plausible because neurodegenerative MCI involves a slow, continuous loss of neuroplasticity, leading to more frequent, small adjustments partially affecting physical activity. In contrast, non-neurodegenerative MCI patients retain motor capacity but have fragile cognitive control. Similarly to what we observed with sleep continuity and duration, increased physiological and cognitive hyperarousal may result in larger, abrupt changes. On the other hand, in the physical activity dimension, the unstable routine mass doesn't highlight differences across cohorts. Finally, comparing cooking habits is more difficult because only one subject in the \textbf{ND} group experienced a change. Nevertheless, the magnitude of change in this dimension is significant, suggesting that shifts in IADLs are often highly pronounced.

\begin{figure}[h!]
    \centering

    \begin{subfigure}{0.45\textwidth}
        \centering
        \includegraphics[width=\linewidth]{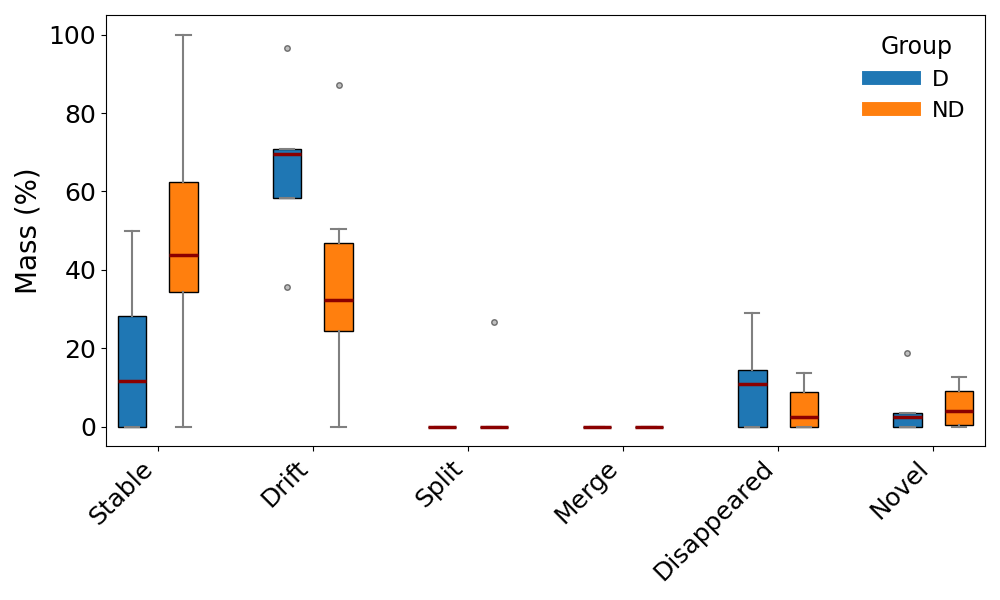}
        \caption{Sleep: Stages}
    \end{subfigure}\hfill
    \begin{subfigure}{0.45\textwidth}
        \centering
        \includegraphics[width=\linewidth]{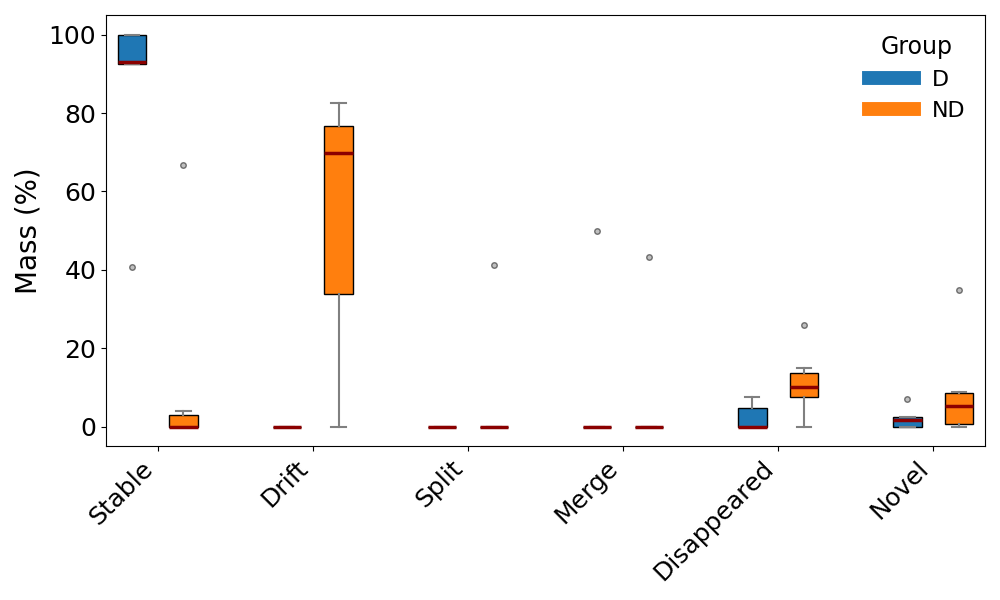}
        \caption{Sleep: Continuity and Duration}
    \end{subfigure}

    \medskip

    \begin{subfigure}{0.45\textwidth}
        \centering
        \includegraphics[width=\linewidth]{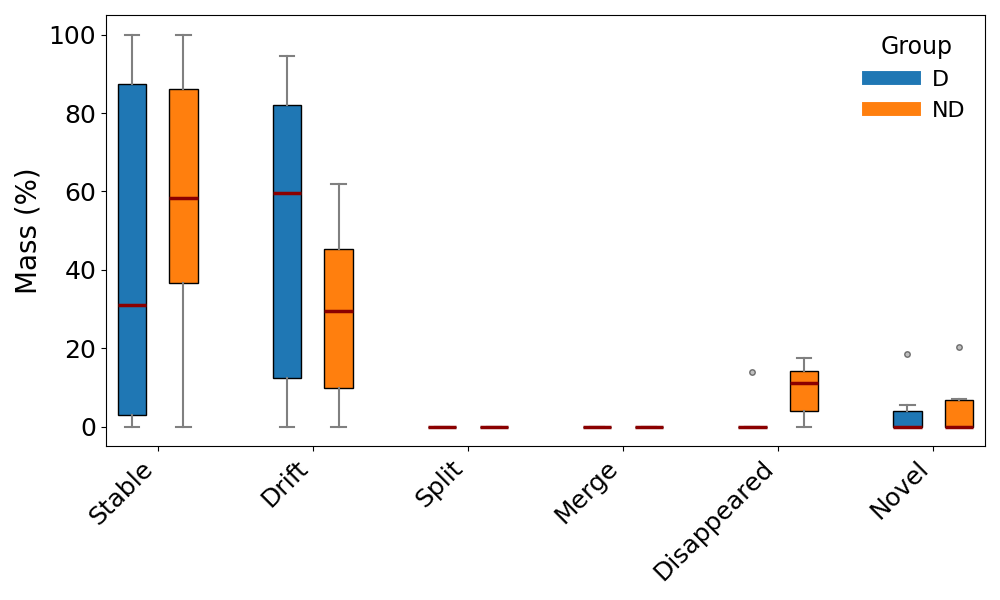}
        \caption{Sleep: Schedule}
    \end{subfigure}\hfill
    \begin{subfigure}{0.45\textwidth}
        \centering
        \includegraphics[width=\linewidth]{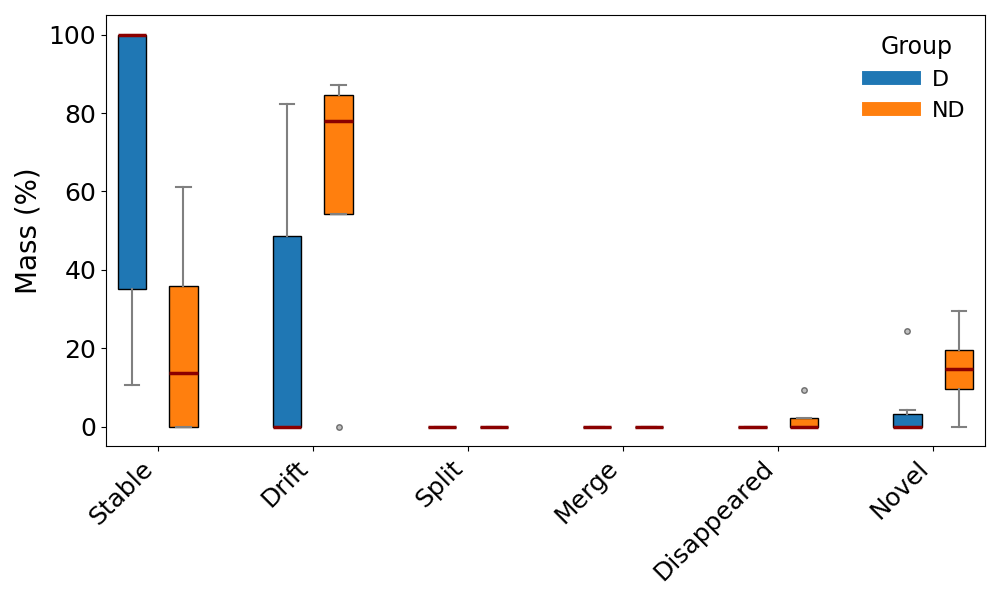}
        \caption{Physical Activity: Steps and Elevation}
    \end{subfigure}

    \medskip

    \begin{subfigure}{0.45\textwidth}
        \centering
        \includegraphics[width=\linewidth]{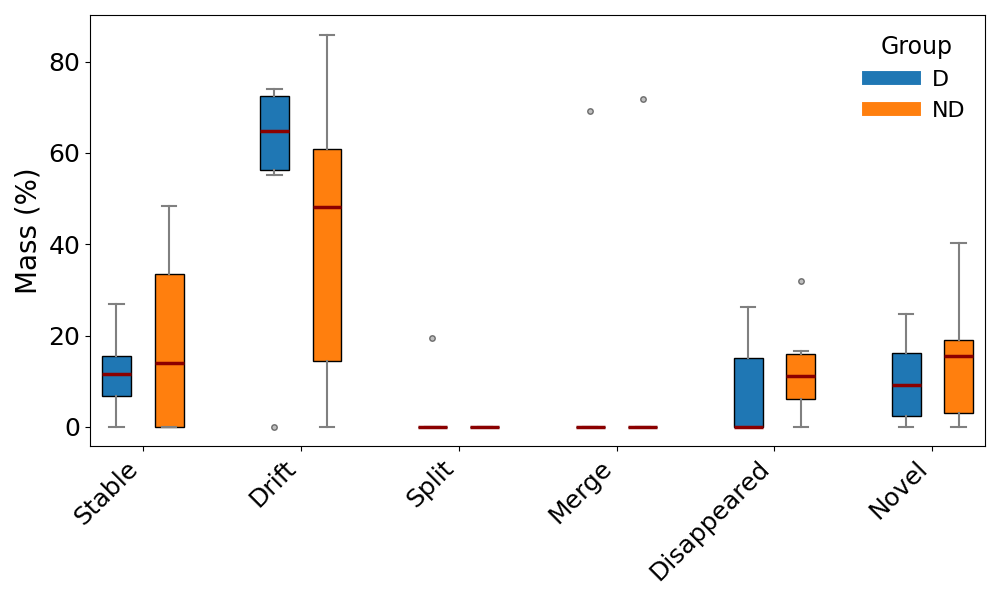}
        \caption{Physical Activity: Intensity}
    \end{subfigure}\hfill
    \begin{subfigure}{0.45\textwidth}
        \centering
        \includegraphics[width=\linewidth]{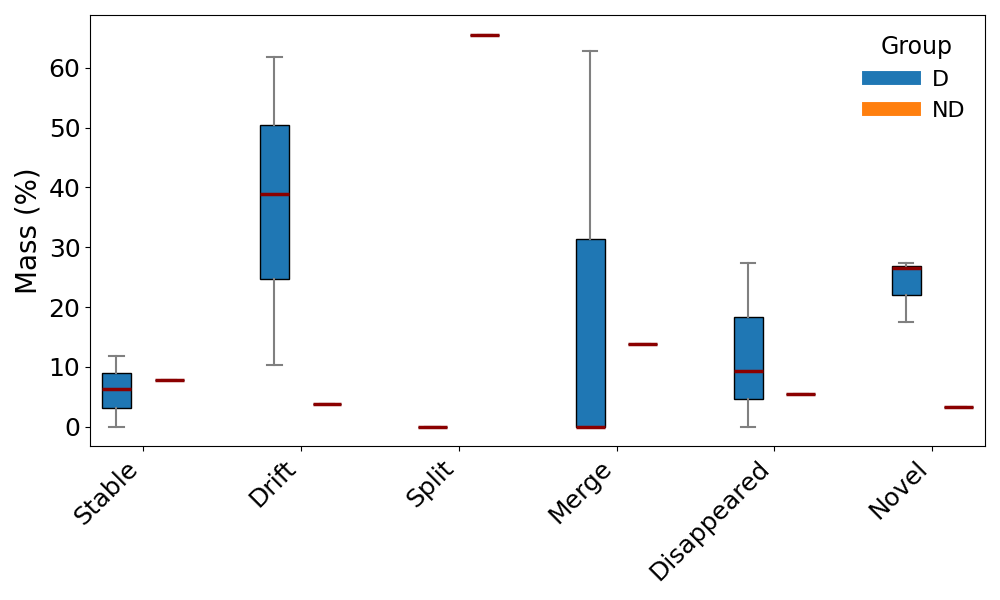}
        \caption{Cooking}
    \end{subfigure}
    
    \caption{Distribution of change types across cohorts}
    \label{fig:evolution}
\end{figure}

\begin{figure}[h!]
    \centering
    \includegraphics[width=0.5\linewidth]{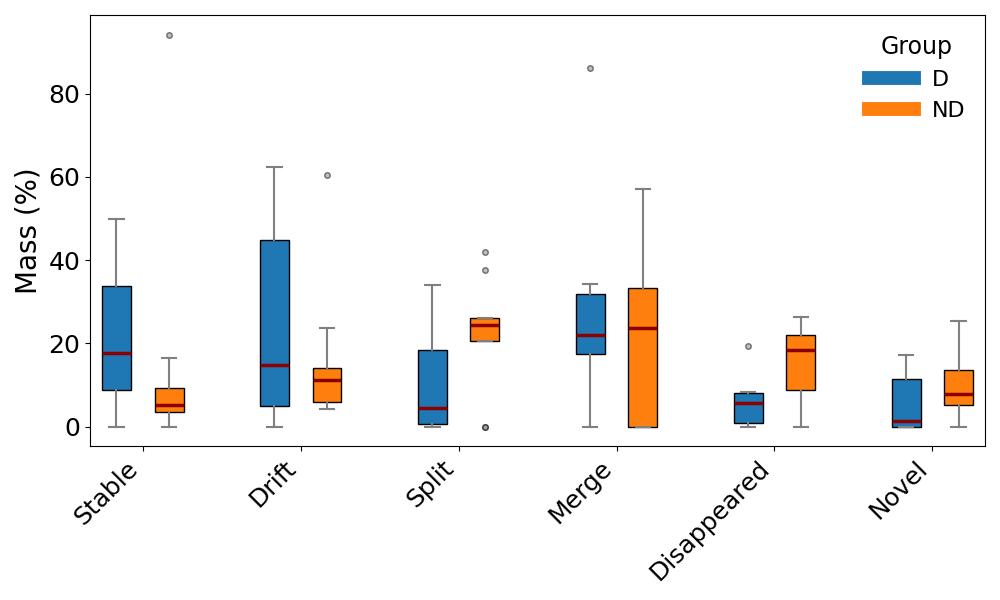}
    \caption{Distribution of change types considering a single comprehensive sleep dimension (stages+schedule+duration\&continuity)}
    \label{fig:global_sleep_evolution}
\end{figure}

Figure~\ref{fig:evolution} reports the average mass associated with each routine evolution type. 
Notably, considering physical activity intensity, we observe that group \textbf{D} is associated with more drifting routines. This insight was not observable in \textit{unstable routine mass} only, which averaged out this information. In general, across dimensions, most of the mass is concentrated in stable and drift events, while disappeared and novel routines are less prevalent. Split and merge events occur relatively rarely. This behavior is expected, as the considered dimensions are intentionally fine-grained and rely on a limited set of coherent features, which favors continuity of routines across segments. To illustrate the effect of representation granularity, Figure~\ref{fig:global_sleep_evolution} shows the distribution of evolution events for a comprehensive sleep dimension obtained by combining sleep stages, duration and continuity, and schedule features. 

In this coarser representation, split and merge events occur more frequently, with a corresponding reduction in stable and drift mass. This arises because changes in the comprehensive dimension may involve only a subset of features, leading to partial similarity across segments. For example, a subject who previously slept longer on weekends may exhibit a merging of weekday and weekend routines if overall sleep duration becomes similar, even if other features (e.g., related to sleep-stage composition) remain almost unchanged. Consistent with this increased heterogeneity, we also observed a larger number of detected change points for both cohorts when using this coarser representation.

\subsection{Sensitivity Analyisis}
\label{sec:sensitivity}

We conducted a sensitivity analysis to assess the robustness of \acronym{} with respect to its main hyperparameters, including the penalty parameter $\beta$ used for change point detection, the similarity threshold $\tau_s$ employed in cluster evolution matching, and the drift threshold $\tau_d$ used to identify significant routine changes. Our goal was not to optimize these parameters for maximal effect, but rather to identify operating regions where the system behavior remains stable under moderate perturbations, thereby reducing sensitivity to specific parameter choices.

\subsubsection{Sensitivity to the Change Point Detection Penalty}

\begin{figure}[h!]
    \centering

    \begin{subfigure}[t]{0.30\textwidth}
        \centering
        \includegraphics[width=\linewidth]{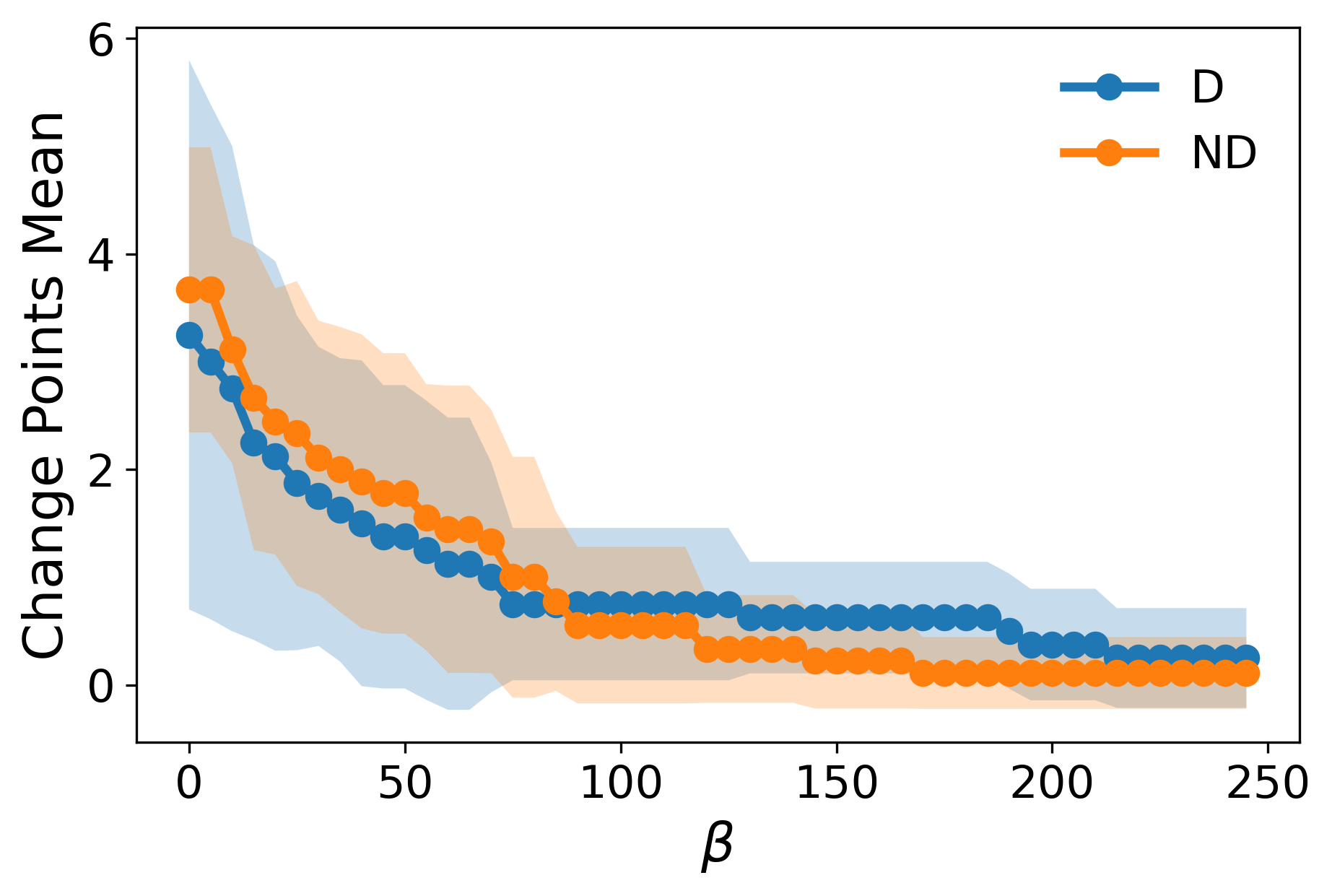}
        \caption{Sleep Stages}
        \label{fig:fig1}
    \end{subfigure}\hfill
    \begin{subfigure}[t]{0.30\textwidth}
        \centering
        \includegraphics[width=\linewidth]{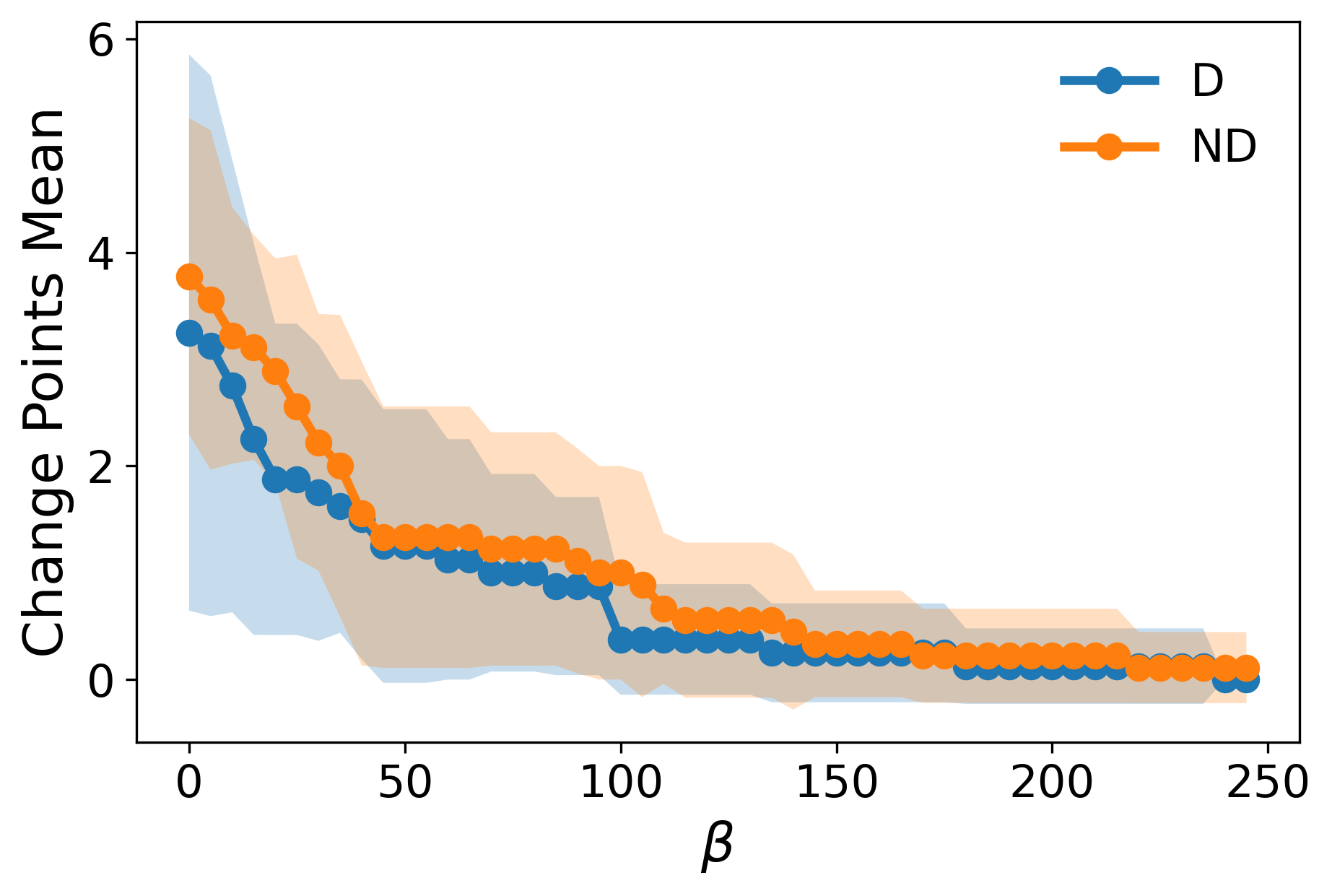}
        \caption{Sleep Continuity and Duration}
        \label{fig:fig2}
    \end{subfigure}\hfill
    \begin{subfigure}[t]{0.30\textwidth}
        \centering
        \includegraphics[width=\linewidth]{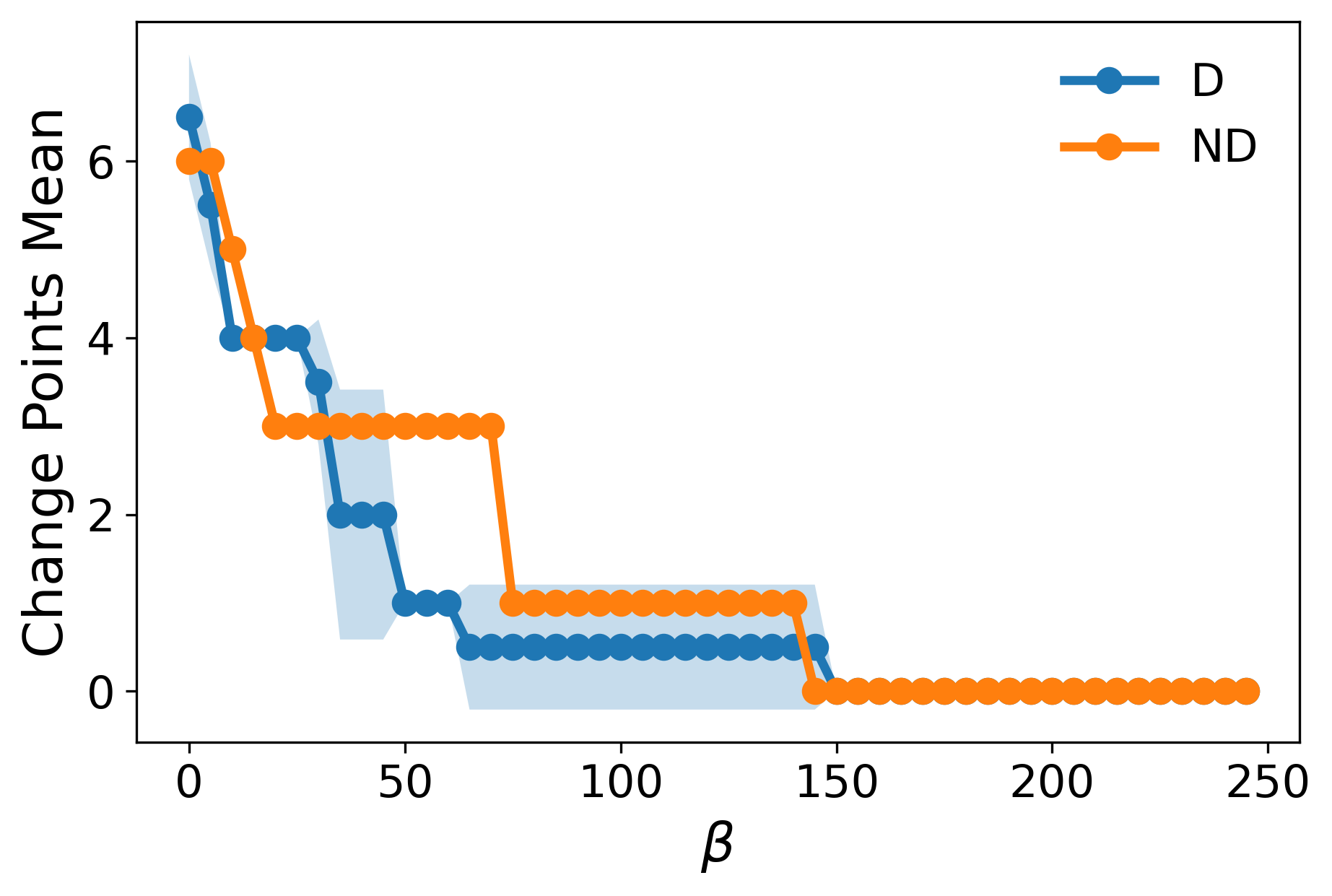}
        \caption{Sleep Schedule}
        \label{fig:fig3}
    \end{subfigure}

    \medskip

    \begin{subfigure}[t]{0.30\textwidth}
        \centering
        \includegraphics[width=\linewidth]{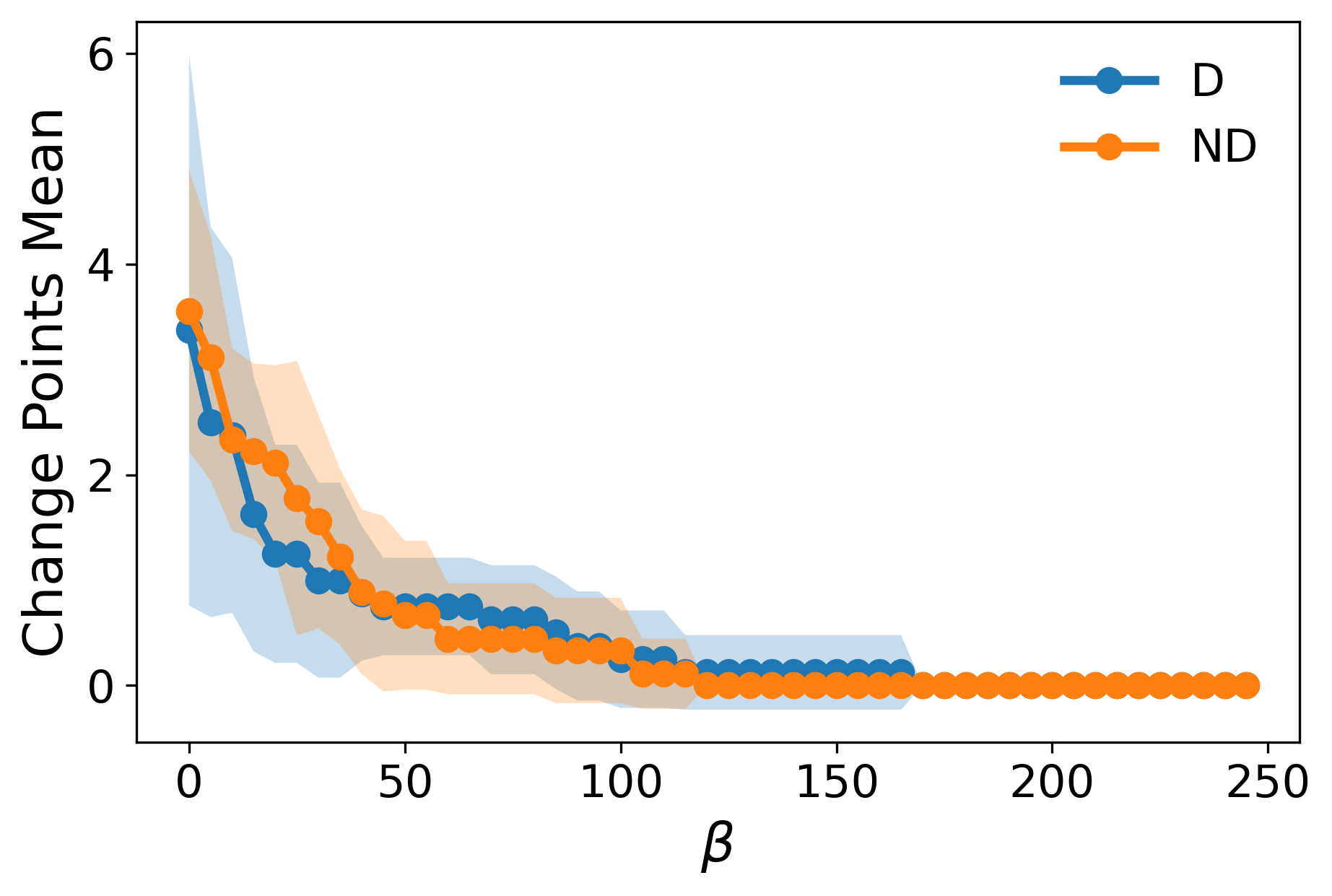}
        \caption{Physical Activity}
        \label{fig:fig4}
    \end{subfigure}\hfill
    \begin{subfigure}[t]{0.30\textwidth}
        \centering
        \includegraphics[width=\linewidth]{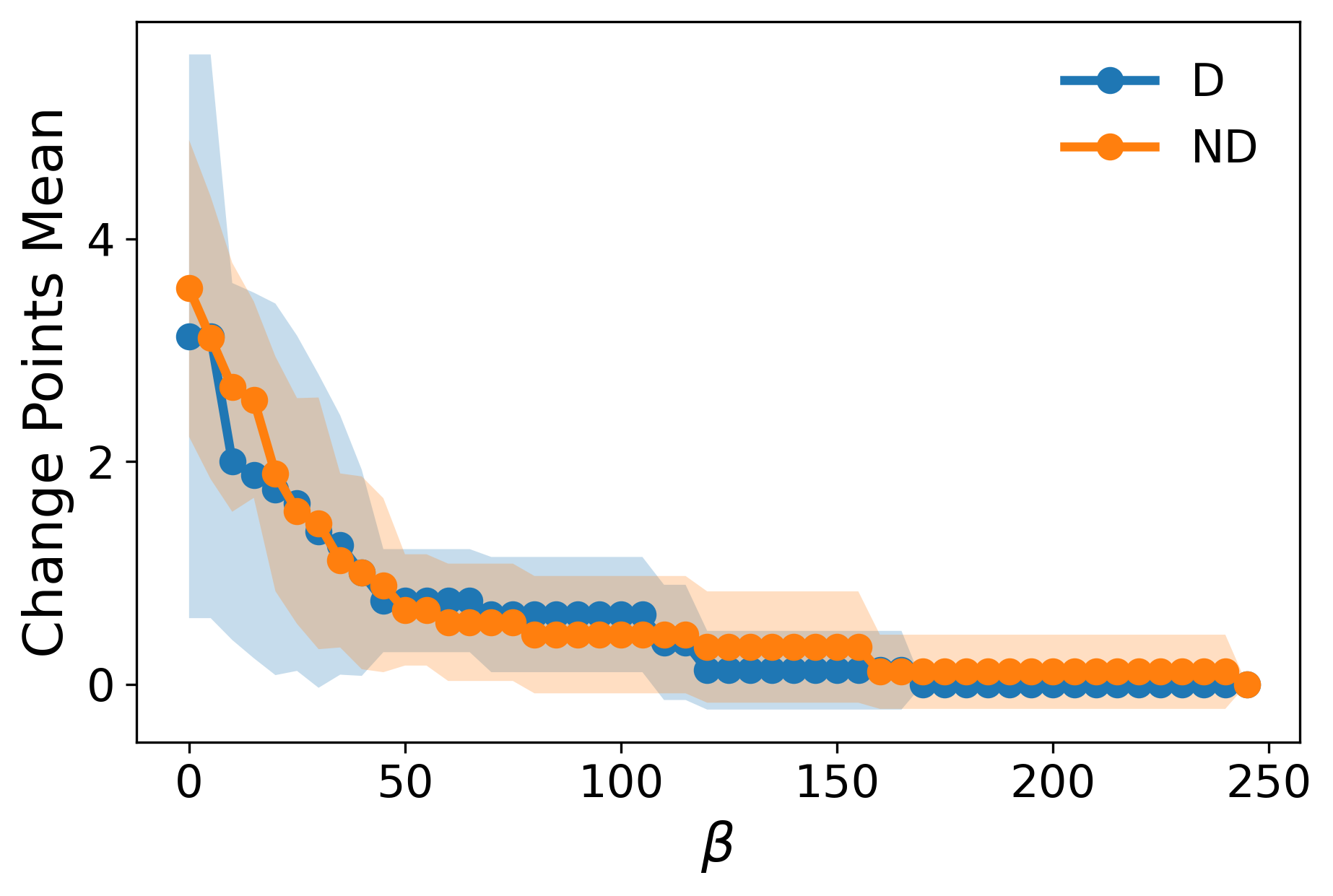}
        \caption{Physical Intensity}
        \label{fig:fig5}
    \end{subfigure}\hfill
    \begin{subfigure}[t]{0.30\textwidth}
        \centering
        \includegraphics[width=\linewidth]{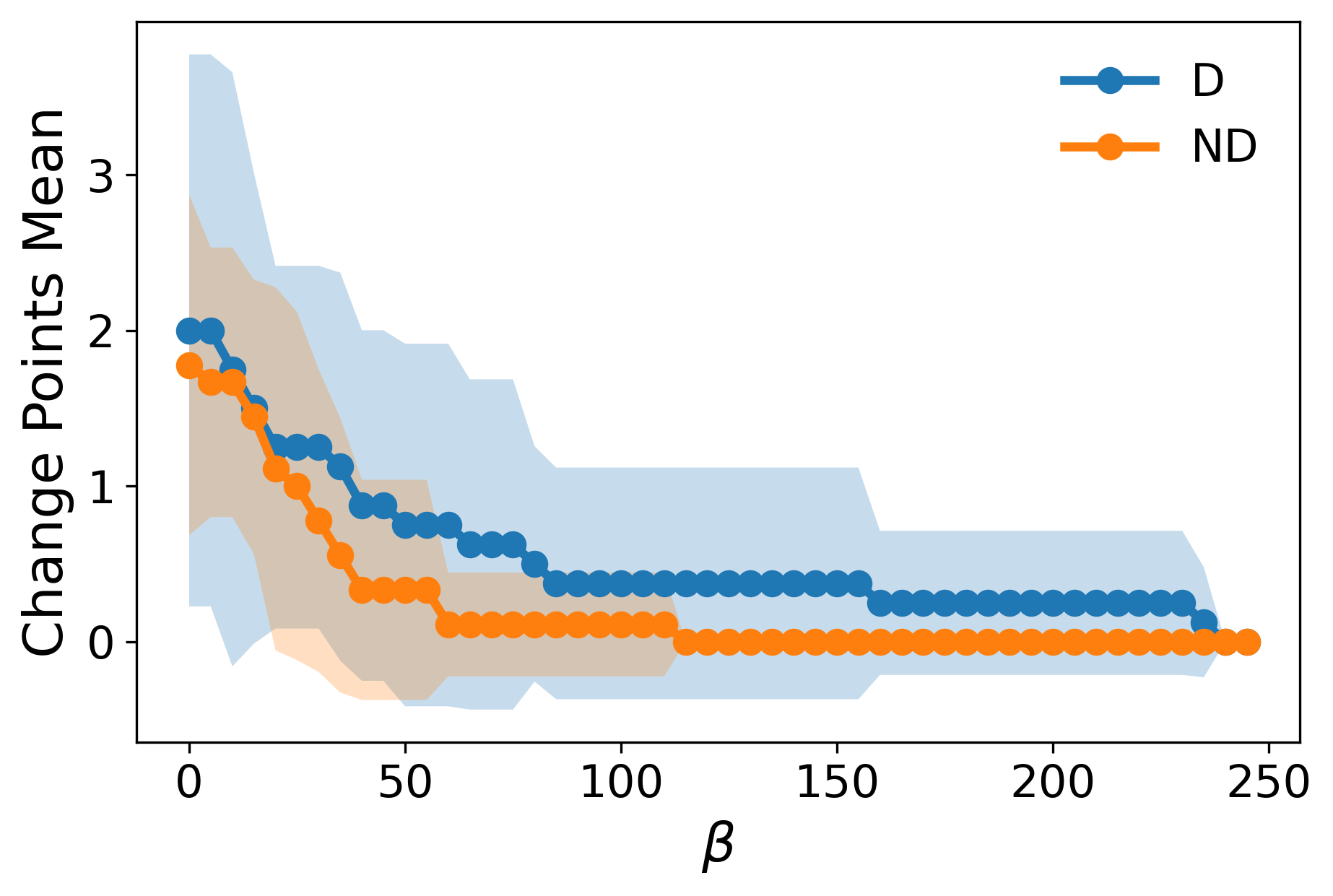}
        \caption{Cooking}
        \label{fig:fig6}
    \end{subfigure}

    \caption{Impact of parameter $\beta$ on the different dimensions across cohorts}
    \label{fig:beta_impact}
\end{figure}

Figure~\ref{fig:beta_impact} illustrates the effect of varying the penalty parameter $\beta$ on the number of detected change points across behavioral dimensions. 
As expected, increasing $\beta$ produces a monotonic decrease in the number of detected change points, reflecting the higher cost assigned to introducing additional segment boundaries. Low values of $\beta$ result in many short-lived segments that likely capture transient fluctuations, whereas very large values lead to over-smoothing and few or no detected changes. Between these regimes, a broad plateau emerges in which the number of detected change points remains stable across dimensions. The selected value $\beta = 60$ lies within this plateau, indicating robustness to moderate parameter perturbations and yielding segments that reflect persistent behavioral changes over clinically meaningful time scales, consistent with the minimum segment duration of three months.

When examining differences across cohorts, we observed that for subjects in group \textbf{D}, non-zero change points in the \textit{Sleep Stages} and \textit{Cooking} dimensions persist even at higher values of $\beta$, whereas they tend to disappear earlier for subjects in group \textbf{ND}. Since larger values of $\beta$ favor only long-lasting and structurally significant changes, this persistence suggests that routine reorganization in these dimensions is more pronounced and sustained for group \textbf{D}. In contrast, changes observed in group \textbf{ND} appear more sensitive to penalization, indicating a higher likelihood of reflecting transient variability rather than enduring shifts in routine structure. While not diagnostic, this pattern suggests that alterations in sleep-stages and instrumental daily activities may be more characteristic of group \textbf{D} at the level of longitudinal routine dynamics.

\subsubsection{Sensitivity to the Similarity and Drift Thresholds}

\begin{figure}[h!]
    \centering

    \begin{subfigure}[t]{0.45\textwidth}
        \centering
        \includegraphics[width=\linewidth]{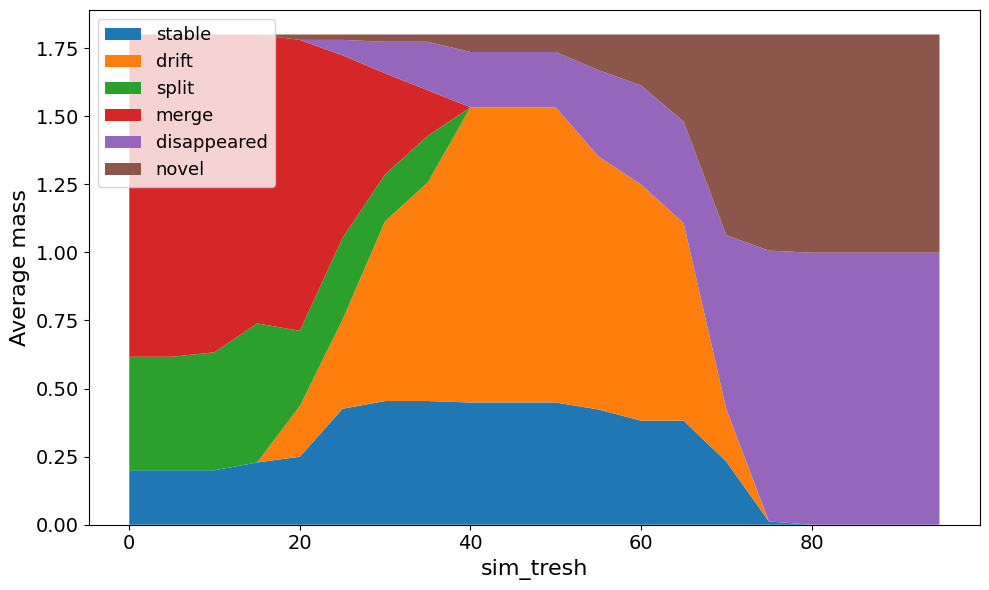}
        \caption{\textbf{D}: how $\tau_s$ impacts events mass ($\tau_d = 0.25)$}
        \label{fig:fig1}
    \end{subfigure}
    \hfill
    \begin{subfigure}[t]{0.45\textwidth}
        \centering
        \includegraphics[width=\linewidth]{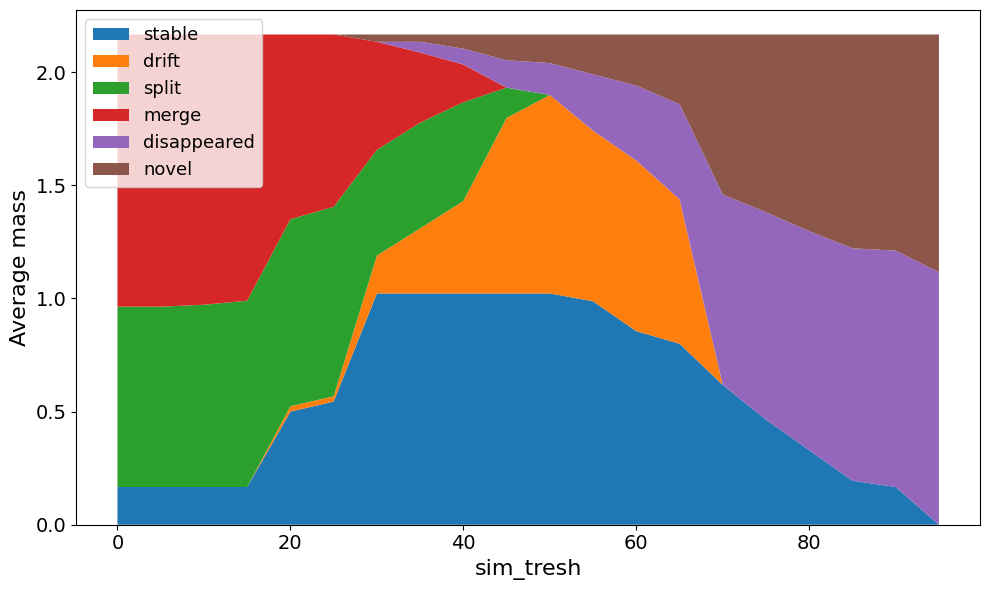}
        \caption{\textbf{ND}: how $\tau_s$  impacts events mass ($\tau_d=0.25$)}
        \label{fig:fig2}
    \end{subfigure}

    \par\medskip

    \begin{subfigure}[t]{0.45\textwidth}
        \centering
        \includegraphics[width=\linewidth]{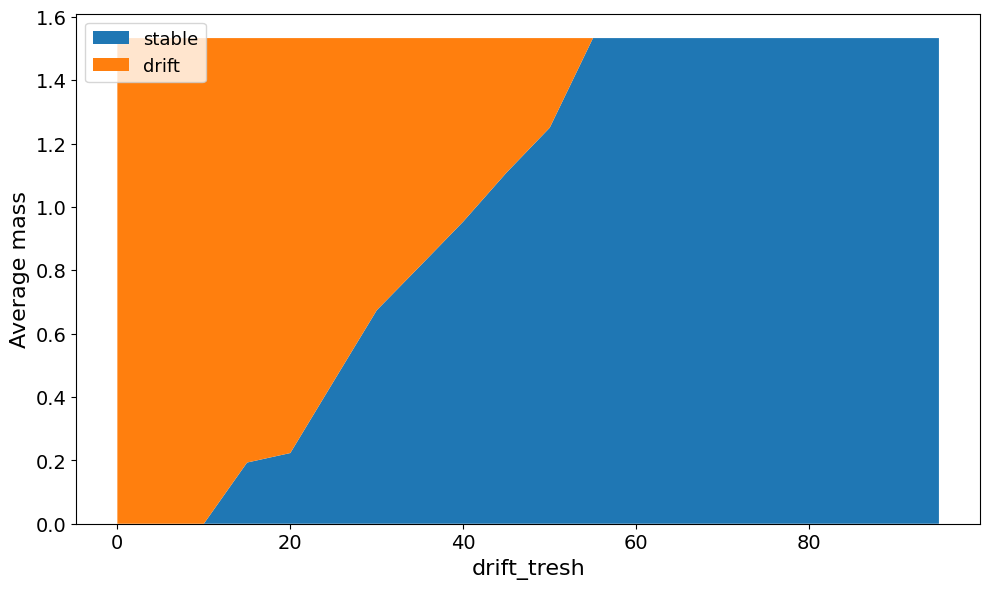}
        \caption{\textbf{D}: how $\tau_d$ impacts detected drifts ($\tau_s = 0.5)$}
        \label{fig:fig4}
    \end{subfigure}
    \hfill
    \begin{subfigure}[t]{0.45\textwidth}
        \centering
        \includegraphics[width=\linewidth]{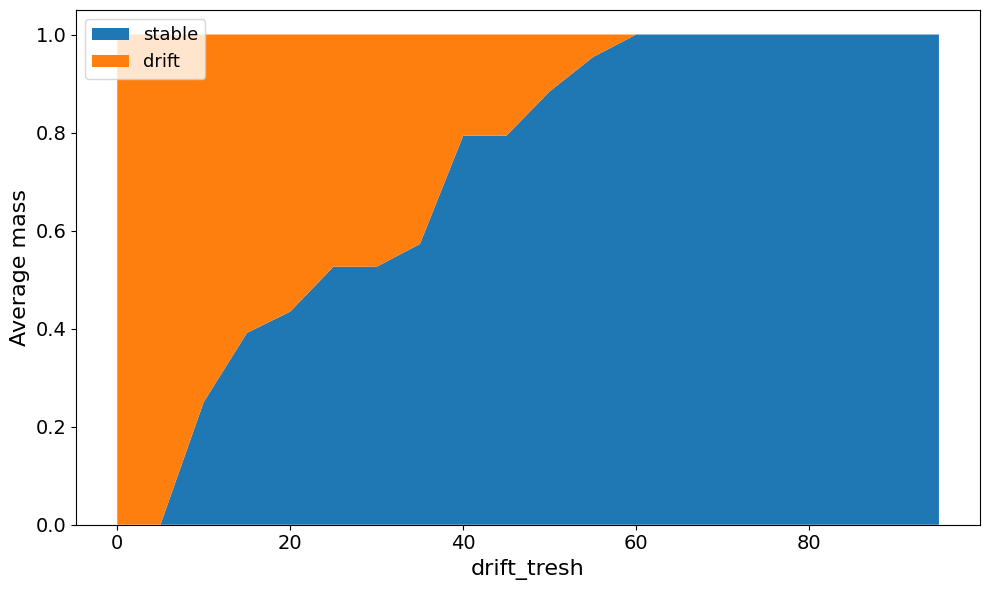}
        \caption{\textbf{ND}: how $\tau_d$ impacts detected drifts ($\tau_s = 0.5)$}
        \label{fig:fig5}
    \end{subfigure}

    \caption{Sleep stages: sensitivity analysis of drift and similarity threshold on cluster evolution tracking.}
    \label{fig:drift_sim_tresh_sleep}
\end{figure}

Figure~\ref{fig:drift_sim_tresh_sleep} shows the effect of varying the similarity threshold $\tau_s$ and drift score threshold $\tau_d$ on the distribution of routine evolution events for the \textbf{D} and \textbf{ND} cohorts. For the sake of space, we show a representative example with the \textit{sleep stages} dimension.
As $\tau_s$ increases, the criterion for matching clusters between RP and OP becomes more stringent. Consequently, the mass of stable routines decreases, while the mass associated with drifted, novel, and disappeared routines increases. On the other hand, when $\tau_s$ is low, cluster matching becomes more permissive, leading to frequent merge and split events, as loosely similar routines across segments are more likely to be associated with one another.

Importantly, within an intermediate range of $\tau_s$ values (approximately $0.4$ to $0.6$), the qualitative distribution of evolution events remains relatively stable. The selected value $\tau_s = 0.5$ lies within this stable region, representing a balance between overly permissive matching (which would mask meaningful routine changes) and overly strict matching (which would fragment routines excessively). 
This suggests that the identified evolution patterns are not artifacts of a particular similarity threshold choice.


\subsubsection{Sensitivity to the Drift Threshold $\tau_d$}

Figure~\ref{fig:drift_sim_tresh_sleep} also illustrates the sensitivity of drift detection to the drift threshold $\tau_d$, with the similarity threshold fixed at $\tau_s = 0.5$.
Lower values of $\tau_d$ result in a larger number of routines being classified as drifted, including minor centroid or variance changes, while higher values progressively restrict drift detection to more substantial behavioral shifts.

We observe in \textbf{ND} a plateau region in which the number and mass of detected drift events remain relatively stable across a range of $\tau_d$ values. The selected value $\tau_d = 0.25$ lies within this region, avoiding both excessive sensitivity to minor fluctuations and under-detection of meaningful routine changes. We observed similar plateau regions in the other dimensions.

Comparing cohorts, we observe a clear difference in how drift responds to increasing $\tau_d$. For group \textbf{D}, the drift mass decreases almost monotonically as $\tau_d$ increases, with a single small plateau region for $\tau_d \in [15,20]$, indicating that drift changes span a wide range of magnitudes and are progressively filtered out under stricter thresholds. In contrast, group \textbf{ND} exhibits multiple plateau regions, suggesting that many drift events have similar strength and remain consistently detected over a range of $\tau_d$ values. This pattern indicates that routine drift in \textbf{ND} is more homogeneous and stable with respect to thresholding, whereas drift in \textbf{D} appears more gradual and heterogeneous.

\subsection{Preliminary Expert Assessment}

To assess the interpretability and semantic coherence of the explanations generated by \textsc{Clusters2Text}, we conducted a limited preliminary assessment with the clinical team collaborating with us in the longitudinal study (i.e., neurologists specialized in cognitive disorders). The goal of this evaluation was not to establish diagnostic or prognostic validity, but to examine whether the explanations produced by \acronym{} are faithful to the underlying data and meaningful within a clinical reasoning context.

The assessment focused on two dimensions:
(1) \textit{Faithfulness}, defined as whether the textual explanation is supported by the corresponding sensor data and observed trends; and
(2) \textit{Interpretability}, defined as the clarity, relevance, and usability of the explanation from a clinical perspective.
We randomly selected from our dataset a 
subset of detected behavioral changes drawn from both the \textbf{D} and \textbf{ND} cohorts, and including different dimensions. 
We provided each description of behavioral change together with the information about the patient it referred to the clinicians. They were also given access to a clinical dashboard, developed as part of the project, that allows them to explore and visualize the longitudinal sensor data of each patient.

Across all reviewed cases, the clinical team confirmed that the described behavioral changes were consistently supported by the visualized sensor data, indicating that the explanations were grounded in the observed data. This suggests that the grounding strategy adopted in \acronym{} effectively constrains the LLM output and mitigates unsupported statements.
Moreover, the clinical team provided a global feedback regarding the quality of the behavior change description provided by \acronym{} referring to the three components of the output:
\begin{itemize}
\item \textbf{Global Behavioral Trend (Part A) and Potential Implications (Part C):} These components were judged to be concise, clear, and useful, offering rapid high-level insight suitable for time-constrained clinical workflows. The language of part C was considered professional and aligned with the language used by neurologists.
\item \textbf{Habit Dynamics (Part B):} This component was identified as overly detailed and technical. While the information was considered accurate, this level of statistical granularity exceeded what is typically required for clinical interpretation and would be more appropriate for technical inspection or system-level analysis.

\end{itemize}

Overall, this assessment suggests that \acronym{} produces explanations that align with clinical reasoning about longitudinal behavioral change. Future work will address the issues raised for Part B, probably adopting a higher level of abstraction and filtering out low-level details to better support practical clinical use.

\section{Discussion}
\label{sec:discussion}

\subsection{Impact}

\acronym{} has potential impact at both the methodological and clinical levels. From a behavioral modeling perspective, it aligns with a growing body of literature showing that cognitive and functional decline often emerges through gradual reorganization of daily routines rather than abrupt failures in isolated activities~\cite{johansson2015cognitive}. Nonetheless, \acronym{} is general and may also apply to other health domains involving long-term monitoring (e.g., rehabilitation, aging-in-place support, and chronic disease management).

An important aspect is that each detected change can be grounded in the underlying data. For instance, for a routine split, a clinician could inspect sensor data associated with the cluster in RP and the two corresponding clusters in OP, understanding which concrete behavioral patterns drove the change.

From a clinical standpoint, \acronym{} is not intended as a diagnostic tool, but as a decision-support system. By identifying sustained behavioral changes that persist over clinically meaningful time scales, the framework could support early risk stratification and pre-alerts for potential transitions toward more vulnerable behavioral profiles. Such early signals may help prioritize follow-up assessments and reduce reliance on invasive or costly clinical procedures (e.g., neuroimaging).

Finally, the explainable-by-design nature of the framework, together with clinician-oriented natural language summaries, facilitates integration into clinical workflows by providing interpretable descriptions of how and when routines change.

\subsection{Towards Online Change Point Detection}
\label{subsec:online}

In its current form, \acronym{} operates in a retrospective setting, analyzing completed observation windows to characterize behavioral change. An important direction for future work is the adaptation of the change point detection component to operate on streaming data, enabling prospective monitoring of behavioral dynamics as new observations become available. In such a setting, change points would serve to flag potential shifts in behavior, supporting earlier human-in-the-loop assessment when appropriate.
Adapting the framework to an online context poses several technical challenges. In particular, normalization strategies and model parameters in the current pipeline rely on access to the full observation period, which is not available in streaming scenarios. Future work will therefore need to explore incremental normalization, adaptive parameter estimation, and mechanisms to ensure the stability and interpretability of detected changes over time. In addition, an online formulation must explicitly address the trade-off between timeliness and reliability, determining how much evidence must be accumulated before a detected shift can be considered sufficiently robust to warrant further analysis.

\subsection{Detected changes are not always clinically relevant}

\acronym{} separates the detection of behavioral changes from their interpretation. The detection components are designed to be sensitive to reorganizations in longitudinal behavioral data, and therefore may surface changes arising not only from clinically meaningful behavioral evolution, but also from contextual factors (e.g., seasonal routines or travel), sensing artifacts, or variations in wearable adherence. 

By grounding explanations in interpretable features, routine-level representations, and explicit cluster evolution events, \acronym{} makes the nature of each detected change more transparent, allowing clinicians to distinguish sustained reorganizations of daily habits from transient or non-behavioral effects, to dismiss detections that are not clinically relevant, and to focus attention on changes that may warrant follow-up. However, we are aware that generated explanations may not always be informative enough to deduce clinical relevance.

In future work, we will explore mechanisms to further support the identification of non-behavioral confounders. One direction is the integration of explicit contextual signals, such as calendar events, known travel periods, or weather information, to annotate detected changes with plausible external explanations. Another direction is the incorporation of data quality and adherence indicators (e.g., sensor uptime, wearable wear-time estimates) directly into the explanatory process, allowing explanations to explicitly flag changes that may be driven by sensing artifacts rather than behavioral reorganization. 
In addition, future versions of \acronym{} could leverage longitudinal consistency checks across multiple behavioral dimensions to distinguish isolated, dimension-specific shifts from coherent cross-domain changes that are more likely to reflect genuine behavioral evolution.

\subsection{LLM hallucinations}

To mitigate LLM hallucinations, \acronym{} decouples behavioral analysis from language generation. Instead of exposing the model to raw sensor logs, multimodal data are first summarized into statistically validated centroids and topological change labels (e.g., \textit{drift}, \textit{disappeared}, \textit{novel}). Centroids are defined over interpretable features for which semantic descriptors are available. The resulting pipeline provides a structurally grounded intermediate representation that constrains the LLM to validated abstractions, preventing the invention of patterns not present in the data. A two-step prompting strategy further reinforces this constraint by first converting centroids into textual routine descriptions and only then requesting an interpretation of behavioral change, under explicit instructions to remain factual and avoid unsupported clinical inferences.  As a result, hallucinations are substantially mitigated. Also, note that \acronym{} uses LLMs fine-tuned in the medical domain, which should further reduce the risks.
Despite these measures, the risk of residual hallucinations remains a significant challenge for clinical reliability, due to the inherent stochastic nature of generative models.  Future work should investigate deterministic fact-checking layers that automatically audit narratives against raw sensor values. 
Moreover, a multi-agent LLM architecture splitting responsibilities across specialized agents (e.g., change interpretation, quantitative verification, and consistency checking) could further improve clinical reliability.

\section{Conclusion and Future Work}
In this work, we presented \acronym{}, an explainable, sensor-based framework for detecting and characterizing behavioral changes in smart home environments. Using longitudinal real-world sensor data collected from individuals with MCI, we showed how behavioral change can be described in terms of routine reorganization. The evaluation demonstrates that \acronym{} provides clinicians with structured and interpretable representations of behavioral changes, enabling the identification of digital markers that may be promising for anticipating cognitive decline due to neurodegenerative pathologies. In particular, within our dataset, changes in sleep stages and cooking-related activities emerge as especially informative for distinguishing subjects belonging to the \textbf{D} and \textbf{ND} cohorts. Moreover, we also received positive feedback from the clinical team regarding the explanations of identified behavioral changes.

%
While the preliminary results are encouraging, the current evaluation does not include a sufficiently large or heterogeneous population. This limitation is primarily due to stringent recruitment constraints, including: (i) strict clinical criteria for mild cognitive impairment (MCI) required by clinicians for enrollment; (ii) structural and logistical constraints (e.g., participants were required to live alone, have reliable network connectivity, and reside within specific geographical areas); and (iii) limited acceptance, as only approximately $30\%$ of eligible candidates agreed to participate. In addition, the substantial infrastructural and long-term maintenance efforts required for sensor deployment and data collection further constrain scalability and participant recruitment. 
Moreover, our evaluation focused only on a subset of behavioral dimensions selected in consultation with clinicians, but the dataset includes additional dimensions that remain unexplored (e.g., outdoor mobility, personal hygiene, etcetera). As a result, a broader experimental validation is required to rigorously assess robustness, generalizability, and sensitivity to both inter-individual and dimension variability. In addition, the proposed approach has not yet undergone a formal clinical validation. Establishing its clinical reliability and relevance will require standardized clinical protocols and systematic involvement of domain experts.

Beyond these limitations and the one discussed in Section~\ref{sec:discussion}, several additional research directions are worth exploring. In this study, we deliberately avoided the use of latent representations, as their stability and interpretability under limited data remain open challenges in clinician-facing, unsupervised settings. Nevertheless, future work will investigate how deep embeddings can be leveraged to identify latent behavioral routines and how such representations can be mapped back to human-readable descriptions. Given the limited amount of data typically available for each subject, one promising direction is the use of federated learning to learn shared representations across multiple individuals while preserving subject-level data isolation.

Another important line of future work concerns the automation of the many hyperparameters required by \acronym{}. In the absence of ground-truth labels, parameter tuning remains particularly challenging. A potential solution is to frame this problem within a Reinforcement Learning paradigm, using a multi-objective reward function that balances clustering stability (e.g., via internal validation metrics such as the Silhouette Score) with clinical utility (e.g., confirmation or dismissal of detected behavioral changes). 

Finally, our experimental analysis considered each behavioral dimension independently, reflecting a deliberate trade-off between modeling completeness and interpretability. Nevertheless, modeling cross-dimensional relationships remains an important direction. Promising extensions include analyzing the temporal alignment of change points across dimensions to identify coordinated or cascading effects, as well as constructing higher-level dependency structures (e.g., causal or influence graphs) on top of the change detection outputs.

\bibliographystyle{acm}
\bibliography{references}

\end{document}